\documentclass[lettersize,journal]{IEEEtran}
\usepackage{amsmath,amsfonts}
\usepackage{algorithmic}
\usepackage{algorithm}
\usepackage{array}
\usepackage[caption=false,font=normalsize,labelfont=sf,textfont=sf]{subfig}
\usepackage{textcomp}
\usepackage{stfloats}
\usepackage{url}
\usepackage{verbatim}
\usepackage{graphicx}
\usepackage{cite}
\usepackage{booktabs}
\usepackage[table]{xcolor}
\usepackage{multirow}

\begin{document}

\title{Orchestrated Vulnerability Management for Heterogeneous Networks: Adaptive Two-Stage Vulnerability Assessment, Context-Aware Risk Prioritization, and Automated Mitigation}


\author{
Ricardo Lopes,
José Moura, 
and Rui Neto Marinheiro%
\thanks{Ricardo Lopes, José Moura and Rui Neto Marinheiro are with ISCTE -- Instituto Universitário de Lisboa, Lisbon, Portugal (e-mails: Ricardo\_Amado\_Lopes@iscte-iul.pt; jose.moura@iscte-iul.pt; rui.marinheiro@iscte-iul.pt).}
\thanks{José Moura and Rui Neto Marinheiro are with Instituto de Telecomunicações, Lisbon, Portugal.}
\thanks{This work was supported by FCT -- Fundação para a Ciência e Tecnologia, I.P., under Project 2024.07624.IACDC (DOI: 10.54499/2024.07624.IACDC). This work was also partially funded by FCT/MECI through national funds and, where applicable, co-funded by EU funds under UID/50008: Instituto de Telecomunicações.}
}


\markboth{Journal of \LaTeX\ Class Files,~Vol.~14, No.~8, August~2021}%
{Shell \MakeLowercase{\textit{et al.}}: Orchestrated Vulnerability Management for Heterogeneous Networks}


\maketitle

\begin{abstract}
Heterogeneous networks introduce significant security challenges due to device diversity, fragile operating conditions, and heterogeneous firmware and service configurations. Traditional vulnerability management approaches typically rely on static scanning and severity-based prioritization, overlooking exploitation likelihood and the operational context of affected assets. Consequently, mitigation actions are often delayed, increasing the exposure window and imposing unnecessary operational overhead on security teams. This paper proposes a SOAR-orchestrated vulnerability management framework that integrates passive asset discovery, an adaptive two-stage vulnerability assessment process, context-aware risk assessment, and automated SDN-based mitigation. The detection engine progressively characterizes the attack surfaces of heterogeneous devices through adaptive assessment strategies tailored to the capabilities of each device, thereby minimizing disruption to resource-constrained IoT assets. Risk assessment combines CVSS severity, EPSS exploitation probability, and contextual attributes to prioritize vulnerabilities according to their actual operational risk Based on the assigned risk band, mitigation is automatically enforced through coordinated OpenFlow and IDS policies, ranging from monitoring and selective service isolation to complete host quarantine. Experimental results demonstrate the effectiveness of the proposed framework. The adaptive two-stage vulnerability assessment reduces scan time by up to 91\%, while identifying 71\% of the baseline vulnerabilities during the initial assessment stage before selectively triggering further analysis. The proposed context-aware risk model reduces by approximately 75\% the number of vulnerabilities requiring immediate mitigation without missing any vulnerability with verified exploitation. Compared with conventional vulnerability assessment, the framework reduces the assessment time for 32 physical hosts by up to 45\% and enforces mitigation within milliseconds, enabling efficient and scalable vulnerability management through adaptive assessment, context-aware prioritization, and automated mitigation.
\end{abstract}


\section{Introduction}
\label{sec:introduction}

Modern networks comprise a diverse set of devices ranging from traditional servers and workstations to embedded sensors, actuators, and industrial controllers \cite{nist23,kokila26,gatti25}, each with distinct hardware capabilities, operating systems, communication protocols, and maintenance cycles \cite{nist23,almasabi24}. While this heterogeneity enables increasingly sophisticated cyber-physical services, it also considerably enlarges the attack surface available to adversaries \cite{sullivan23}. The increasing volume of disclosed vulnerabilities and the uneven distribution of real-world exploitation make proactive vulnerability management on network devices essential to be further investigated \cite{jacobs21,shimizu26,10599429}. \par

Traditional vulnerability management solutions generally follow a sequential workflow consisting of asset discovery \cite{11205517}, vulnerability scanning \cite{ecik21}, risk prioritization \cite{shimizu26}, and mitigation \cite{11205517}. Although widely adopted, these approaches present several limitations when applied to heterogeneous IoT-enabled networks \cite{kokila26}. Device discovery and attack surface mapping commonly rely on active scanning techniques that generate additional traffic and may degrade the performance of resource-constrained devices or disrupt their operation \cite{ecik21}. Exhaustive active scanning can also require considerable execution time, limiting its applicability in dynamic environments where devices frequently join, leave, or change their network behavior \cite{ecik21,sullivan23}.

Risk prioritization remains a significant challenge in vulnerability management \cite{varvarigou25}. Existing approaches commonly rely on the Common Vulnerability Scoring System (CVSS) to prioritize vulnerabilities, assuming that higher severity scores correspond to higher operational risk \cite{allodi14}. However, several studies \cite{shimizu26,mell2006common} have shown that CVSS severity alone is a poor predictor of real-world exploitation, as many high-severity vulnerabilities are never exploited, whereas lower-severity vulnerabilities may become attractive targets due to the availability of public exploits or ongoing attack campaigns \cite{shimizu26}. Moreover, CVSS base score does not account for the security posture of the target environment, such as the presence of firewalls, network segmentation, access control mechanisms, or other compensating security controls that can significantly reduce the effective risk posed by a vulnerability \cite{mell2006common}. Consequently, effective vulnerability management requires integrating additional factors, including exploitation likelihood \cite{jacobs21,allodi14}, asset criticality \cite{varvarigou25}, network exposure \cite{gatti25}, the operational role of each device \cite{gatti25,adewole25}, and endorsing in real-time the most-critical cyber risks by proportionate mitigation decisions \cite{varvarigou25}.

Mitigation strategies adopted by conventional vulnerability management systems also remain predominantly coarse-grained \cite{miettinen17}. Actions such as complete host isolation or network quarantine effectively reduce exposure but may unnecessarily interrupt legitimate services, particularly in IoT environments where devices frequently support safety-critical or business-critical operations \cite{nist23}. As heterogeneous networks continue to grow in scale and complexity, there is an increasing need for adaptive mitigation mechanisms capable of balancing cybersecurity requirements with operational continuity \cite{varvarigou25}.

Recent advances in Software-Defined Networking (SDN) \cite{kokila26} and Security Orchestration, Automation and Response (SOAR) \cite{kinyua2021ai} technologies provide a promising foundation for addressing the challenges discussed in the previous paragraphs. By separating the control and data planes, SDN enables centralized and programmable traffic management, allowing security policies to be dynamically enforced at different granularities \cite{kokila26,11205517}. Complementarily, SOAR platforms automate security workflows by coordinating asset discovery, vulnerability assessment, threat intelligence, alert correlation, and automated response across multiple security components \cite{kinyua2021ai}. Together, SDN and SOAR can offer proactive vulnerability analysis and automatic mitigation \cite{10599429}, contributing to a continuous adaptation of protection strategies to network dynamics \cite{11205517}.

Motivated by the above mentioned limitations and oportunities, the current work is guided by the following research questions (RQs):

\begin{itemize}
    \item \textbf{RQ1:} How can heterogeneous assets and their attack surfaces be rapidly and accurately identified using hybrid passive-active detection with adaptive fast/deep scanning, while minimizing operational disruption to resource-constrained devices?

    \item \textbf{RQ2:} How can vulnerability prioritization be improved by integrating CVSS severity, exploitation likelihood, and operational context into a dynamic risk assessment?

    \item \textbf{RQ3:} How can SOAR and SDN technologies be integrated to automate fine-grained, risk-driven mitigation while minimizing operational disruption and preserving service continuity?
    
    \item \textbf{RQ4:} How does the proposed SDN- and SOAR-based vulnerability management framework affect the performance and scalability of heterogeneous programmable networks?
\end{itemize}

\begin{figure*}[!t]
\centering
\includegraphics[width=\textwidth]{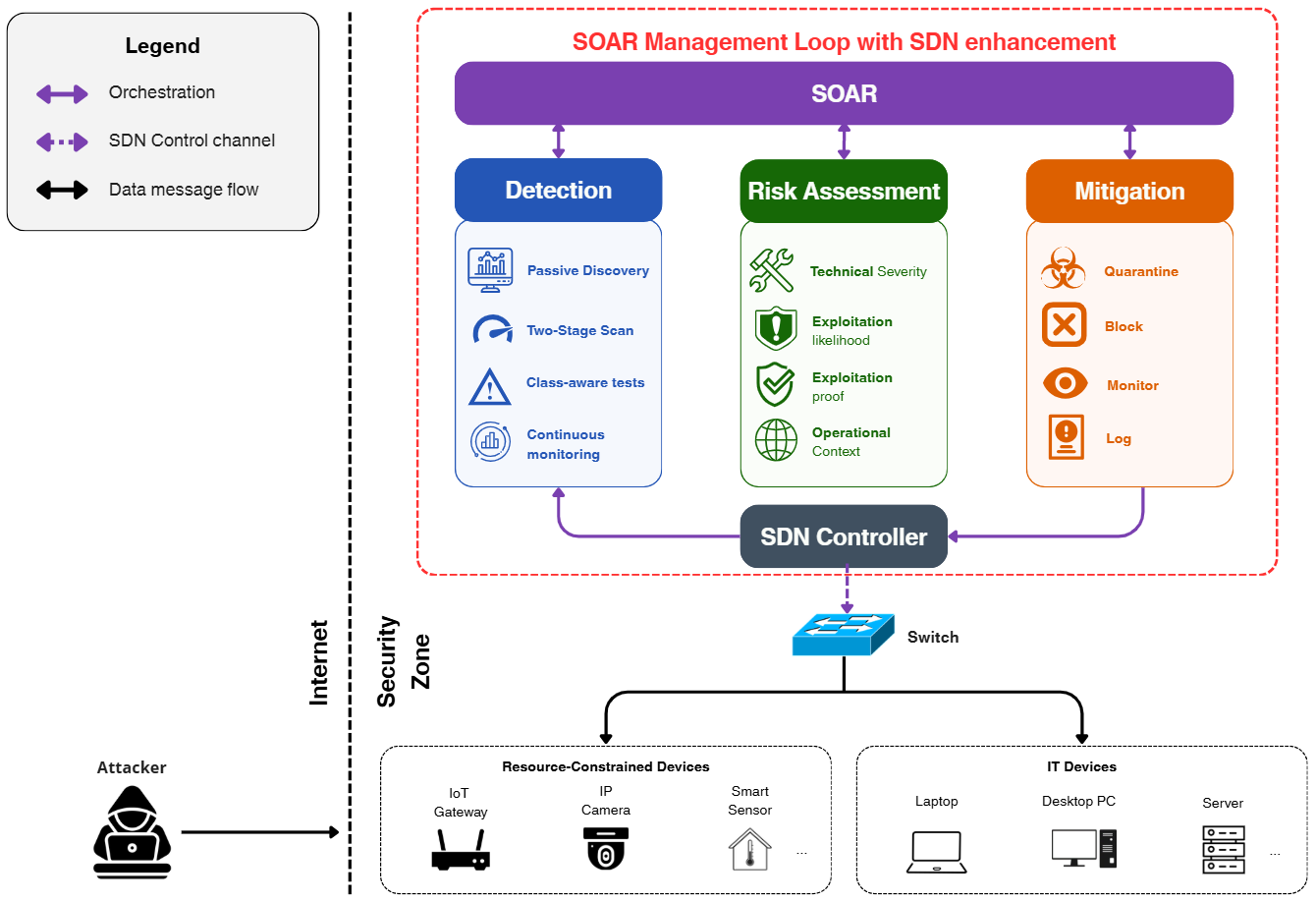}
\caption{Proposed SOAR-driven vulnerability management framework with SDN enforcement.}
\label{fig:Fig1}
\end{figure*}

To answer these research questions, the current paper proposes an adaptive, risk-driven vulnerability management framework that combines passive asset discovery, adaptive two-stage vulnerability assessment, dynamic risk assessment, automated security orchestration, and programmable mitigation, as shown in Figure~\ref{fig:Fig1}. At the core of the proposed architecture is a SOAR orchestrator responsible for coordinating the complete vulnerability management lifecycle, from asset discovery and vulnerability assessment to risk evaluation and mitigation deployment. This orchestration layer enables the framework to automatically correlate information collected from multiple security components, determine the most appropriate response according to the evaluated risk \cite{varvarigou25}, and consistently enforce mitigation policies throughout the network. To accommodate dynamic and heterogeneous network environments, the framework continuously discovers and monitors devices that may be automatically instantiated \cite{s23052762,nguyen2020horizontal} and services that may be dynamically reallocated \cite{zavodovski2019exec} in response to dynamic service demand. Consequently, newly deployed resources are promptly incorporated into the vulnerability management lifecycle.

The proposed detection engine continuously observes network traffic to passively identify assets and characterize their attack surfaces, while selectively performing active probing only on devices capable of safely supporting deeper inspection. Vulnerability prioritization combines CVSS severity, Exploit Prediction Scoring System (EPSS) probabilities, and contextual information regarding device criticality, exposure, and operational role to estimate the cyber risk of each vulnerability. The SOAR orchestrator continuously correlates this information and coordinates the execution of mitigation workflows. According to the assessed risk, it interacts with the SDN controller and the Intrusion Detection System (IDS) to enforce proportional responses ranging from enhanced monitoring and event logging to selective service isolation or complete device isolation, thereby minimizing operational disruption while maintaining network security.

The paper main contributions are as follows:

\begin{itemize}

\item The design of a SOAR orchestration framework that continuously integrates adaptive vulnerability discovery, context-aware risk assessment, and automated mitigation into a unified end-to-end vulnerability management workflow (Subsection~\ref{subsec:design_goals_architecture}).

\item The design of an adaptive two-stage vulnerability assessment architecture, preceded by passive asset discovery, that progressively characterizes the attack surfaces of heterogeneous network devices while minimizing assessment overhead on resource-constrained IoT assets (Subsection~\ref{subsec:design_detection}).

\item The proposal of a context-aware vulnerability risk assessment model that combines CVSS severity, EPSS exploitation probability, and operational context to prioritize vulnerabilities according to their actual operational risk, substantially reducing the number of vulnerabilities requiring immediate mitigation without sacrificing protection against vulnerabilities with verified exploitation (Subsection~\ref{subsec:design_assessment}).

\item The implementation of a fine-grained risk-driven mitigation strategy that coordinates the SDN controller and IDS to dynamically enforce traffic mirroring and monitoring, service isolation, or complete host quarantine according to the assessed vulnerability risk (Subsection~\ref{subsec:impl_mitigation}).

\item The experimental validation of the deployed framework on virtualized and hybrid testbeds (Section~\ref{sec:evaluation}), yielding the following results: (i) up to a 91\% reduction in adaptive scan time while preserving 71\% vulnerability coverage after the fast scan (Subsection~\ref{sec:constrained_aware_scanning}); (ii) a 75\% reduction in vulnerabilities requiring immediate mitigation without missing verified exploited vulnerabilities (Subsection~\ref{sec:exploitation_aware_risk}); (iii) millisecond-scale mitigation enforcement (Subsection~\ref{sec:pipeline_latency}); and (iv) up to a 45\% lower execution time for the concurrent assessment of 32 hosts compared with a representative OpenVAS-based vulnerability assessment approach \cite{11205517} (Subsection~\ref{sec:scalability}).

\end{itemize}

The remainder of this paper is organized as follows. Section~\ref{sec:literature_review} reviews the state of the art in vulnerability management, analyzes the limitations of existing approaches, and identifies the research gaps addressed by the proposed framework. Section~\ref{sec:design} presents the architecture and design of the proposed framework. Section~\ref{sec:implementation} details its implementation. Section~\ref{sec:evaluation} evaluates the framework through a set of experimental scenarios and discusses the obtained results. Section~\ref{sec:discussion} balances the implications and limitations of the proposed framework. Finally, Section~\ref{sec:conclusions} concludes the paper and outlines future research directions.

\section{Literature Review}
\label{sec:literature_review}

\noindent This section reviews the literature relevant to the proposed framework. Subsection~\ref{subsec:vulnerability_detection} discusses existing vulnerability detection approaches. Subsection~\ref{subsec:vulnerability_risk_assess} reviews methods for vulnerability risk assessment and prioritization, while Subsection~\ref{subsec:vulnerability_mitigation} examines mitigation strategies for vulnerable assets. Finally, Subsection~\ref{subsec:literature_limitations} summarizes the limitations of existing solutions and identifies the research gaps addressed by the proposed framework.

\subsection{Vulnerability Detection}
\label{subsec:vulnerability_detection}

\noindent The proliferation of IoT devices within heterogeneous access networks introduces a significant problem for vulnerability detection, as there is a need to establish a balance between depth of inspection and system stability. To effectively manage this problem, network security practices have traditionally relied on two fundamental paradigms, namely passive monitoring and active scanning, both of which have significant advantages and limitations. Passive monitoring involves the continuous observation of non-dataplane traffic, such as, DHCP \cite{kumar2019all} or DNS \cite{campo2024inferring,guo2020detecting} normal messages, without the injection of any external network packets, ensuring that the detection process does not become intrusive and maintaining system stability. In contrast, active scanning involves the direct interrogation of network endpoints, where extra injected packets are sent to the target network endpoints to identify any misconfigurations, open ports, and unpatched software. \par

Recent advancements in passive detection methodologies have seen the increasing adoption of artificial intelligence techniques to process network traffic with minimal overhead. For example, modern network architectures are now integrating fog computing with ML classifiers to infer potential vulnerabilities in the network through the patterns of network traffic, offloading the processing overhead from IoT devices \cite{almasabi24}. This non-intrusive detection has been further enhanced in programmable network environments, where modern frameworks are now integrating deep learning classifiers \cite{kokila26} and ML-driven Intrusion Detection Systems (IDS) \cite{sahbi23} in the SDN control plane. This enables the continuous analysis of flow statistics to classify abnormal network activities and potential threats without the need to introduce active probes into the network. However, the critical disadvantage of purely passive methodologies is their inherent inability to detect potential threats, which is due to their fundamental requirement to engage in active network communication. A comparative analysis of different methodologies of discovering vulnerabilities in the system has revealed that passive monitoring systems are inherently incapable of monitoring services that do not actively transmit data over the network and, therefore, do not account for dormant vulnerable services in the system until they are actively targeted and exploited by an adversary in the system \cite{ecik21}. \par

To achieve comprehensive visibility and uncover dormant services, active scanning is required. However, indiscriminate probing can interfere with resource-constrained environments \cite{ecik21}. These environments typically have limited TCP/IP stacks and constrained computational resources, making them susceptible to resource exhaustion and unintended Denial of Service (DoS) conditions when exposed to traditional network scanning techniques \cite{nist23}. \par

To mitigate these issues of operational risk, recent models make use of a sequential hybrid approach, conducting a passive network profiling process before undertaking an active process of network scanning \cite{miettinen17}. Information gathered in passively regarding device types based on admission data and behavioral signatures, enables a more accurate approach in the active scanning process by configuring probes based on the identified target's capacity to maximize scanning while maintaining network stability \cite{sullivan23}. \par

Beyond simple device profiling, modern detection frameworks are shifting toward semantic, context-informed inference on the detected vulnerabilities \cite{huang2026aegisguard}. In addition, to address the high false-positive rates of traditional scan tools, recent architectures like AegisGuard \cite{huang2026aegisguard} utilize Large Language Models (LLMs) and Retrieval-Augmented Generation (RAG) to process structured host telemetry, such as, running services, privilege states, and kernel configurations. By combining this live telemetry with external threat intelligence, these systems can detect configuration-dependent and environment-specific vulnerabilities that version-based matching alone fails to capture. This approach allows for a "grey-box" analysis that inspects internal system states to assess exploit reachability within a specific operational context \cite{barrere2013vulnerability}. \par

The complexity of modern networks necessitates that detection result in actionable intelligence rather than just dense reporting \cite{farris2018vulcon}. Advanced methodologies now integrate logical reasoning with deep learning to better associate detected vulnerabilities with their likelihood of exploitation \cite{zeng2021licality}. For instance, the LICALITY system employs a neuro-symbolic model to capture attacker preferences and learn threat attributes from real-world attack scenarios, refining initial assessments through probabilistic logic. Similarly, the machine-learning based Vulnerability Priority Scoring System (VPSS) enhances detection by quantifying organizational context, such as the relevance of the network segment, the presence of high-value assets, and the level of protection on hosts \cite{hore2022towards}. \par

To manage the high volume of alerts generated by monitoring engines, the Mosemboli framework \cite{11395359} uses an ensemble of auto-encoders and calibrated uncertainty quantification to prioritize anomaly alerts, ensuring that analysts focus on the most certain threats first.

\subsection{Vulnerability Risk Assessment}
\label{subsec:vulnerability_risk_assess}

\noindent Vulnerability mitigation is commonly prioritized using the Common Vulnerability Scoring System (CVSS). This system primarily measures the intrinsic technical severity of a vulnerability. Although the specification includes temporal and environmental metrics to capture changing circumstances and organisation-specific conditions, these dimensions are often unavailable or inconsistently populated in practice, leaving the Base score as the dominant metric in CVSS evaluations. Furthermore, studies of real-world exploitation show that many vulnerabilities with high CVSS scores are not exploited at scale, while a comparatively small subset accounts for a disproportionate share of observed attacks \cite{allodi14,shimizu26}. As a result, prioritization workflows centered on static severity alone can lead to alert overload and vulnerability fatigue, allocating resources to issues that are severe but not actively exploited.

To incorporate threat intelligence into prioritization, the Exploit Prediction Scoring System (EPSS) was introduced as a data-driven complement to CVSS. The probability that a given vulnerability will be exploited in the near term is estimated and updated regularly by EPSS using statistical learning models \cite{jacobs21}. Recent frameworks extend this dual perspective through vulnerability management chaining, which is a sequential, decision-tree-based approach that first filters vulnerabilities by their active threat level (e.g., using EPSS thresholds and Known Exploited Vulnerability catalogs) and then by their intrinsic CVSS severity \cite{shimizu26}. By setting a strict probability threshold for the severity assessment, this chained methodology drastically reduces the mitigation workload by successfully isolating the small subset of flaws that are both highly severe and actively targeted \cite{shimizu26}.However, recent literature highlights that this dual perspective is incomplete as it only considers vulnerability properties and external threat factors. It does not fully capture organisation-specific exposure and impact, which are determined by the affected asset and its deployment environment \cite{gatti25,varvarigou25}. The authors in \cite{jiang2025survey} confirm the lack of an open-source context-sensitive standard for prioritized risk assessment.  \par

Consequently, translating a vulnerability into actual cyber risk requires evaluating its specific operational environment \cite{abbas25,huang2026aegisguard,barrere2013vulnerability}. Effective prioritization must therefore augment standard vulnerability metrics with structural contextual variables, such as asset criticality, network exposure, and topological dependencies \cite{abbas25,adewole25,hore2022towards}. For example, a medium-severity vulnerability on a core gateway inherently poses a significantly higher systemic risk than a high-severity flaw on an isolated sensor, given the former broadest attack surface and potential for lateral network propagation. \par

Moreover, these contextual variables should change over time. As assets change configuration, activate new services or alter communication behaviour, contextual signals must be continuously monitored to update risk prioritisation. For instance, if a dormant IoT device starts to expose a privileged service or open a sensitive port, its exposure and potential impact will increase, necessitating a revision to its prioritisation to reflect the expanded attack surface \cite{varvarigou25,gatti25}. Overall, contemporary approaches emphasise the need to integrate technical severity (CVSS), exploitation likelihood signals (e.g. EPSS) and dynamic operational context as complementary aspects of a robust risk self-assessment model \cite{huang2026aegisguard}.

Finally, to manage the resource constraints inherent in mitigating discovered threats, detection reports are increasingly integrated into multiobjective optimization frameworks. Systems like VULCON \cite{farris2018vulcon} utilize mixed-integer programming (MIP) to prioritize mitigation based on a vulnerability’s age, persistence, and its impact on the total vulnerability exposure of the organization. Similarly, OPISPA-V \cite{sonmez2020decision} suggests a method that uses Analytical Hierarchical Process and MIP to identify the optimal set of preventative actions for human decision-makers to maximize risk prevention for a fixed financial investment. \par

\subsection{Vulnerability Mitigation}
\label{subsec:vulnerability_mitigation}

\noindent Software-defined networking (SDN) provides a promising foundation for mitigation by decoupling the control plane from the data plane, enabling centralized and programmable network management \cite{kokila26}. This architecture allows for the dynamic reconfiguration of routing rules, which can be utilized to quickly isolate compromised devices or redirect traffic for deep packet inspection \cite{nist23}. Systems like IoT SENTINEL leverage SDN to automatically confine the traffic flows of vulnerable devices when they are inducted into a network, ensuring that potentially insecure "brownfield" legacy devices do not endanger the rest of the infrastructure \cite{miettinen17}. To enhance resilience, SDN-based gateways can divide a network into virtual overlays, such as trusted and untrusted networks, where vulnerable assets are placed in restricted isolation \cite{11205517,miettinen17}. Furthermore, mitigation can be achieved through network segmentation using virtual local area networks (VLANs) or demilitarized zones (DMZs), which act as enforcement boundaries to control access to sensitive components \cite{nist23}. Boundary devices like industrial-class stateful firewalls further support this by enforcing deny-all, permit-by-exception policies between adjacent network levels\cite{nist23}.

Mitigation strategies can be categorized into host-level and flow-level approaches, each balancing security requirements with operational continuity \cite{miettinen17}. Host-level mitigation typically involves endpoint hardening, the use of physical access controls, or the complete disconnection of a target node from the network \cite{almasabi24}. While effective at stopping local activity, complete host isolation can unnecessarily interrupt legitimate services, particularly in IoT environments supporting safety-critical operations \cite{miettinen17}. In contrast, flow-level mitigation offers finer granularity by allowing the system to filter traffic based on specific protocols, IP addresses, or individual flows \cite{almasabi24}. This adaptive approach enables the suppression of malicious traffic (such as a UDP flood or SQL injection attempt) without disabling the device entirely, thus maintaining service availability \cite{varvarigou25}. Research suggests that effective mitigation is best supported by a holistic model that combines low-level platform health indicators (host-level symptoms like CPU and memory anomalies) with network-based assessments of deployed services \cite{gatti25}.

Complementing local mitigation strategies, the Antidose scheme \cite{simpson2018inter} facilitates inter-domain collaboration to remedy volumetric DDoS attacks by using proof-of-work challenges and flow cookies to build verifiable whitelists in upstream Autonomous Systems.

SOAR mechanisms bridge the gap between detection and mitigation by automating complex security workflows \cite{varvarigou25,shimizu26}. Orchestration is frequently applied in virtualized environments where machine learning algorithms integrated into management systems can trigger self-healing responses, such as automatically scaling or rerouting network functions in response to detected faults \cite{varvarigou25}. Automated frameworks can prioritize vulnerabilities into distinct tiers, such as "Critical Priority", which may trigger emergency change procedures and immediate mitigation playbooks \cite{shimizu26}. Some systems also employ deep learning orchestration to ensure secure data communication and integrity across multi-layer IoT ecosystems \cite{kokila26}. While specialized SOAR platforms offer high-end automation, the logic for vulnerability mitigation chaining can often be implemented through simpler scripts or database queries, allowing organizations to strategically focus resources on the small percentage of vulnerabilities that pose an actual exploitation threat \cite{shimizu26}. This automated detection-classification-response cycle significantly reduces the mean time to recovery and minimizes the need for manual intervention \cite{varvarigou25}.

\subsection{Literature Limitations}
\label{subsec:literature_limitations}

\begin{table*}[t]
\centering
\caption{Summary of underexplored research areas motivating the proposed framework.}
\label{tab:underexplored}
\renewcommand{\arraystretch}{1.2}
\begin{tabular}{p{1.7cm} p{7.0cm} p{7.0cm}}
\hline
\textbf{Aspect} & \textbf{Limitation of Existing Approaches} & \textbf{Underexplored Area} \\
\hline

Detection &
Active scanning is intrusive, time-consuming, and may affect IoT availability. &
Need for non-intrusive, continuous asset discovery and attack-surface mapping with minimal network overhead. \\

Risk Analysis &
Risk prioritization combines CVSS and EPSS but ignores dynamic asset context. &
Need to integrate dynamic operational context that is continuously updated as asset behaviour and configuration evolve. \\

Mitigation &
Mitigation commonly consists of coarse host-level isolation. &
Need for adaptive, fine-grained, context-aware mitigation that preserves service availability while reducing risk. \\

\hline
\end{tabular}
\end{table*}

\noindent The authors in \cite{10599429} concluded that the vast majority of previous work manage security vulnerabilities when they are being exploited by threats by reactively mitigating those vulnerabilities. They argue that a new alternative methodology is urgently needed for automatically detecting and proactively mitigating vulnerabilities before they could be exploited. This perspective is followed by the current work.

Traditional literature and security practices often rely on active vulnerability scanning for both device discovery and attack surface mapping \cite{sullivan23}. These active methods are frequently criticized for their intrusive nature, which can generate excessive traffic and lead to unwanted side effects, such as network disruptions or the total failure of resource-constrained assets, such as, some IoT devices \cite{ecik21,nist23}. Furthermore, active detection requires long processing times to exhaustively enumerate installed applications and open services \cite{sullivan23}. Most critically, these broad active scans can overwhelm the limited compute resources of sensitive IoT assets, leading to the degradation or total disruption of their essential functions. To address these shortcomings, the proposed framework enhances existing solutions by implementing passive detection \cite{ecik21,sullivan23} for both devices and their attack surfaces. By continuously monitoring standard network traffic, specifically leveraging protocols, such as, DHCP \cite{kumar2019all} and IPv6 Neighbor Discovery Protocol (NDP) \cite{narten2007neighbor}, the system identifies and characterizes new assets without introducing network overload or interfering with their operational state. This non-intrusive approach provides real-time visibility into asset composition and behavioral characteristics, such as accessed and exposed services. To optimize even more the new proposal efficiency, it begins with a fast scan to quickly conclude initial device identification and attack surface mapping. Following this, the system only performs an additional deep scan for specific devices identified as having sufficient computational resources to support more intensive investigation, thereby ensuring that sensitive or resource-constrained devices remain protected from operational disruption \cite{sullivan23,nist23}. \par

Existing vulnerability-management approaches often rely on severity-oriented risk classification, with CVSS commonly used as the main criterion for prioritizing mitigation. However, severity alone does not reliably represent the probability that a vulnerability will be exploited in practice, since high-severity vulnerabilities are not always the most likely to be weaponized or observed in the wild \cite{allodi14}. To address this limitation, exploit-aware models such as EPSS introduce data-driven estimates of exploitation likelihood, improving prioritization by incorporating threat activity rather than relying exclusively on static technical severity \cite{jacobs21}. More recent work has also emphasized that risk assessment should account for operational and contextual factors, including asset criticality, exposure, and the role of the affected device within the network \cite{shimizu26,gatti25,abbas25}. Building on these insights, the proposed framework classifies risk not only from CVSS severity, but also from exploitation probability and device context, enabling more adaptive and operationally meaningful prioritization in heterogeneous environments. \par

Existing mitigation strategies in vulnerability-management systems often adopt coarse-grained responses, such as host quarantine, VLAN reassignment, or full device isolation, to contain vulnerable or compromised assets \cite{miettinen17}. Although these actions can reduce exposure, they may also introduce significant operational disruption, especially when the affected device provides legitimate services that remain necessary for other parts of the network. In heterogeneous and IoT-enabled environments, this limitation is particularly relevant because devices may support critical or resource-constrained functions, making indiscriminate isolation an unsuitable default response \cite{nist23}. Recent SDN-based approaches show that programmable networks can support more flexible enforcement by applying traffic-control decisions at the flow or service level rather than only at the host level \cite{nist23,sahbi23,kokila26}. Building on this capability, the proposed framework adjusts mitigation to the actual risk and operational context of each finding \cite{nist23,miettinen17}: high-risk cases may trigger full device isolation, while narrower threats can be handled through service-level isolation, enhanced monitoring, or detailed logging of threat-related information. This enables a more proportional response that reduces exposure while preserving service continuity whenever possible. Table~\ref{tab:underexplored} summarizes the less explored literature research topics that motivate the current work. This work main novelty, according to our best knowledge, lies in its closed-loop SOAR orchestration, integrating the entire vulnerability-management lifecycle. Unlike related previous works \cite{11395359,zeng2021licality,sonmez2020decision,simpson2018inter}, which address isolated stages, the proposed framework combines adaptive two-stage assessment for heterogeneous IoT assets, context- and exploitability-aware risk assessment, and fine-grained SDN mitigation, enabling proportional responses (e.g., service isolation) to be enforced autonomously within milliseconds.\par

The next Section debates the design of proposed system.

\section{System Design}
\label{sec:design}

\noindent This Section discusses the design of proposed framework, addressing the limitations identified in Subsection~\ref{subsec:literature_limitations}. First, Subsection~\ref{subsec:design_goals_architecture} defines the design goals and provides a high-level overview of the framework's closed-loop architecture for the automated discovery, assessment, and mitigation of high-risk device security vulnerabilities. The remaining subsections detail the three functional stages of the control loop. Subsection~\ref{subsec:design_detection} presents the first stage, describing how network devices are profiled and safely inspected for security vulnerabilities. Subsection~\ref{subsec:design_assessment} describes the second stage, explaining how detected vulnerabilities are evaluated and prioritized according to their risk against the normal operation. Finally, Subsection~\ref{subsec:design_mitigation} presents the third stage, describing how proportional mitigation actions are automatically enforced to address the highest-risk vulnerabilities.

\subsection{Design Goals \& Architecture Overview}
\label{subsec:design_goals_architecture}

\noindent To establish a proactive security posture, the architectural design of the proposed framework is guided by five foundational principles. These principles ensure scalable threat mitigation while avoiding operational disruptions to the underlying infrastructure. The design principles are as follows:

\begin{itemize}
  \item \textit{Continuous and Event-Driven Monitoring}: Rather than relying solely on predefined scanning schedules, the framework should be able to respond autonomously to real-time network events, such as new devices joining the network and dynamic behavioural changes.

  \item \textit{Safe and Adaptive Inspection}: The system must be able to adapt its vulnerability analysis dynamically to the capabilities and resilience of the target device. This ensures that fragile, resource-constrained nodes can be safely inspected without becoming overloaded or disrupted by aggressive active probing.

  \item \textit{Context-Aware Risk Prioritization}: Threat evaluation must go beyond static severity metrics by incorporating topological context and real-time vulnerability exploitation. This should encompass predictive likelihood models and confirmed threats to calculate actionable risk scores in real time.

  \item \textit{Autonomous and Fine-Grained Mitigation}: The architecture must leverage the programmability of SDN to orchestrate proportional defensive responses without human intervention. Moving away from traditional binary enforcement, the system should be able to dynamically scale its mitigation strategies according to the criticality of the vulnerability detected, effectively neutralising critical threats while maximising legitimate infrastructure service continuity.

  \item \textit{Scalability and Interoperability}: The closed-loop system must seamlessly integrate heterogeneous  devices across orchestration, monitoring, scan, and control roles to ensure rapid threat response, avoiding performance bottlenecks as network density and device diversity increase.
\end{itemize}

\begin{figure}[hbt!]
\centering
\includegraphics[width=\columnwidth]{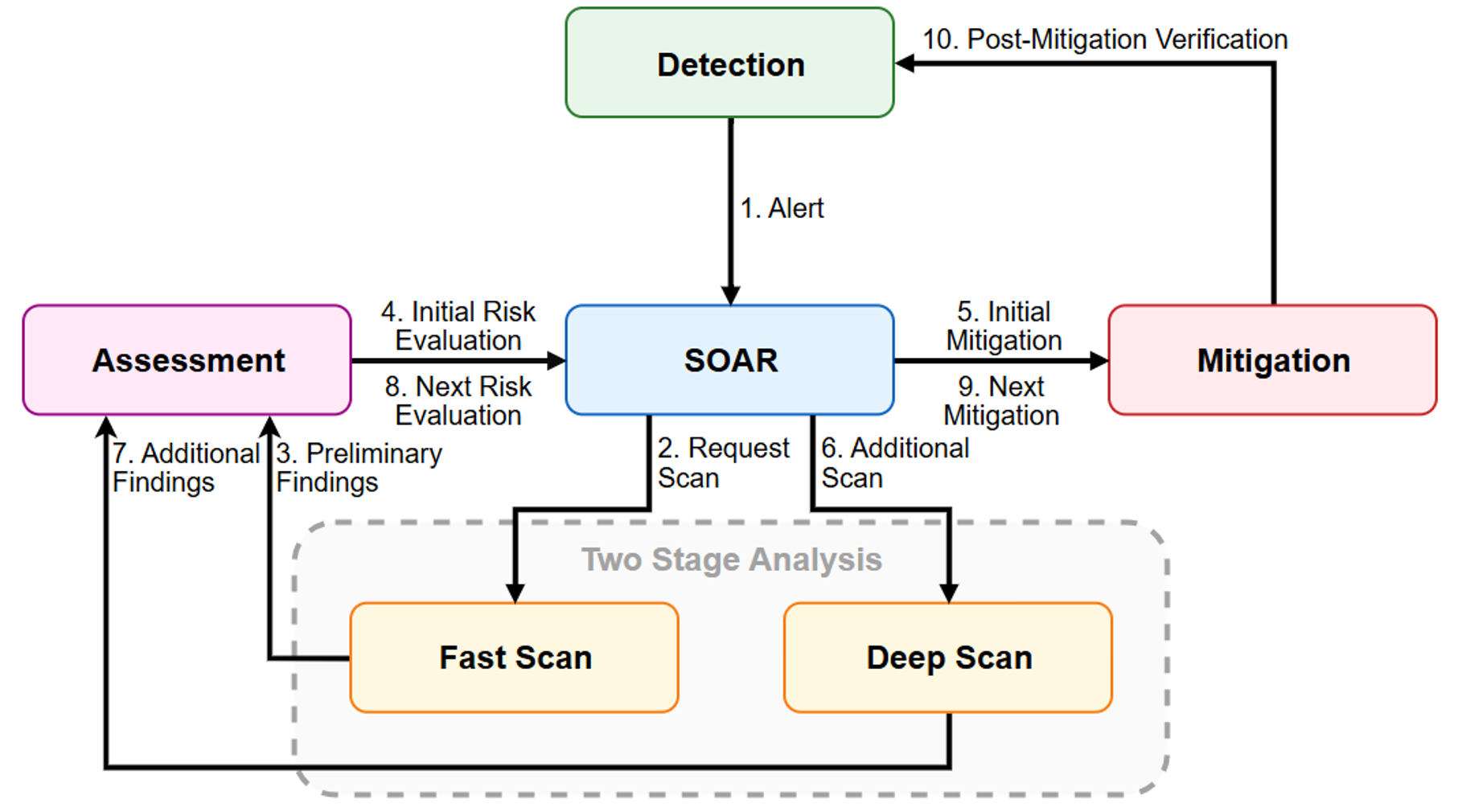}
\caption{SOAR-orchestrated workflow of the proposed framework}
\label{fig:hl_flow}
\end{figure}

To operationalise these principles, the framework adopts the closed-loop architecture shown in Figure~\ref{fig:hl_flow}. The framework has a central orchestrator that coordinates every device through detection, risk assessment, and mitigation before returning the device to continuous oversight. Rather than relying on a single management trigger, the system can trigger security evaluations through three complementary mechanisms: (i) initial network admission; (ii) scheduled rescans tailored to the device class; and (iii) event-driven re-evaluation, which is made whenever a known device exposes a previously unseen port or exhibits anomalous traffic behavior. These complementary inspection triggers enable the orchestrator to coordinate the appropriate response in an evolutionary manner, from the initial detection of a potential threat to the enforcement of the final management decision. Each time the orchestrator is triggered, it enriches its knowledge of the infrastructure's security posture, enabling progressively more informed management decisions and improving its ability to mitigate in real-time high-risk threats.\par

The cycle begins when \textit{Detection} (1) signals the orchestrator to evaluate a device. If the alert is a new device joining the network, the orchestrator first classifies the device by type so that all subsequent inspections can be tailored accordingly. For devices already known to the system, this classification is reused rather than recomputed. The orchestrator then dispatches a \textit{Fast Scan} (2), which is a quick inspection of the most relevant ports whose preliminary findings are forwarded to \textit {Assessment} (3). The assessment evaluates these findings and returns an initial risk evaluation to the orchestrator (4), which sends it to \textit{Mitigation} (5) to enforce a proportional response and minimize exposure as soon as possible. This is also the decision point: if the device has a critical vulnerability, the device is quickly isolated, and the management loop stops here with no further scanning. \par  

Otherwise, the orchestrator dispatches a \textit{Deep Scan} (6), which is a broader inspection of the remaining port range that uncovers latent vulnerabilities. The deep scan forwards its findings to \textit{Assessment} (7), which returns a next risk evaluation to the orchestrator (8). This evaluation is passed to \textit{Mitigation} (9) again, refining the device's protection. \par

Finally, the loop ends, but each device is continuously reevaluated through ongoing monitoring and post-mitigation verification (10). In this way, as schedules elapse, behavior shifts, or new exposure appears, each previous management decision could be revisited as new conditions arise. The management loop is detailed in the next subsections. \par

\subsection{Detection}
\label{subsec:design_detection}
\noindent The proposed detection methodology combines passive device profiling with an adaptive two-stage vulnerability assessment. Unlike conventional vulnerability assessment, which applies the same inspection strategy to every asset, device profiling determines how each asset is assessed, while the assessment dynamically selects the appropriate inspection depth and and techniques according to device characteristics and the outcome of the initial assessment. \par

The methodology begins with near-zero-overhead device profiling, which uses only admission-time metadata such as the hardware vendor identity, the network stack's parameter request signature, and any vendor-declared class information. As these attributes can be observed without active probing, the system can passively infer a device's operating system family and stack characteristics. Each device is then classified as either \textit{resource-constrained} or \textit{traditional IT} based on its profile. If a device cannot be identified with confidence, it defaults to the resource-constrained class. This reflects a deliberate trade-off: a robust device that is handled as constrained receives a lighter scan. However, a fragile device that is handled as robust risks disruption or failure under aggressive probing. \par

To prevent resource exhaustion and unintentional DoS conditions, active inspection is managed by an adaptive, risk-aware probing mechanism. Instead of executing an exhaustive test suite, the system preselects vulnerability tests based on a risk criterion and the device's classification, reducing scan scope and duration by focusing inspection on the most relevant vulnerabilities. For resource-constrained devices, the system relaxes detection-confidence requirements and excludes disruptive test classes in favor of safe, reconnaissance-oriented checks. \par

The inspection scope is further controlled through dynamic port selection within a two-stage scanning strategy. The two assessment stages are deliberately separated to minimise the exposure window between detection and mitigation. Rather than performing a single scan of the full port range, which would delay results until completion, the system first conducts a rapid scan using a dynamically reduced set of historically targeted ports. As this phase examines fewer ports, it delivers actionable results sooner, enabling an earlier defensive response. If vulnerabilities are detected, mitigation is triggered immediately. A finding with the highest severity halts the process before the deep scan, as the device is removed from normal operation and no further inspection is required. Otherwise, the device proceeds to a comprehensive scan of the remaining ports to uncover latent or less common vulnerabilities. By prioritising the fast phase, the system proactively addresses the most likely threats, incurring the cost of exhaustive scanning only after the initial exposure has been reduced.\par

All device classes follow the same two-stage process. The difference is that resource-constrained devices remain under continuous passive monitoring once their scan cycle is complete. As these devices are only inspected with safe, reconnaissance-oriented tests, the scan provides a less complete overview of their state than it does for traditional IT systems. Passive monitoring helps to close this gap by observing network behaviour without injecting traffic and sustaining visibility of the device without the load that deeper active inspection would impose. It produces severity-ranked alerts without placing any load on the device itself, giving ongoing oversight of the most fragile assets while preserving the stability of the heterogeneous environment.

\subsection{Risk Assessment}
\label{subsec:design_assessment}

\noindent The assessment stage transforms raw scan results into a prioritized risk ranking. Relying on technical severity alone can lead to the over-prioritization of vulnerabilities that are serious in theory but rarely exploited in practice. Instead, the assessment first evaluates each finding along two axes, severity and exploitability, and then refines the result according to the device's operational context. Severity measures the potential impact of a successful attack, while exploitability reflects the likelihood of an attack based on either observed or predicted exploitation. These two axes are then combined to place each finding into one of four ordered risk bands. \par

A finding is considered exploited when exploitation is observed or predicted to be likely exploited. When no exploitability information is available, the worst-case scenario is assumed, and the finding is treated as exploited. The decision bands are defined as follows: \par

\begin{itemize}
  \item \textit{CRITICAL}: High severity and exploited.
  \item \textit{HIGH}: Moderate severity and exploited.
  \item \textit{MEDIUM}: High severity, yet not exploited. This is a real but latent risk that is serious enough to track, but not to act on immediately.
  \item \textit{LOW}: Low severity, whose limited impact does not justify intervention regardless of exploitation.
\end{itemize}

The key differentiator is exploitation: a finding that is already being attacked is an immediate concern, so a moderately severe but exploited finding (\textit{HIGH}) outranks a more severe but unexploited one (\textit{MEDIUM}). Additionally, each vulnerability keeps its own band rather than being absorbed into a single verdict for the device. This preserves a per-service view of where the risk lies, enabling later enforcement to act on each affected service in proportion to its own risk rather than penalizing the entire device. \par

Finally, the band is adjusted to reflect the operational context of the device, since the same vulnerability does not carry the same risk everywhere. A flaw is far more dangerous on a highly exposed or critical device than on an isolated one that is rarely reached, and context elevates a finding's band accordingly. One such case is resource-constrained devices: because they are lightly inspected, less is known about their true state. Additionally, because they are inherently more fragile, a vulnerability on one of these devices is more likely to lead to compromise or disruption. These two reasons point toward greater caution, so findings on these devices are elevated above their baseline band. The result is a ranking that reflects not just what a vulnerability is but also where it sits in the network. \par

\subsection{Mitigation}
\label{subsec:design_mitigation}

\noindent The mitigation stage enforces a response that is proportional to the risk band of each finding. Instead of reacting to every vulnerability with full device isolation, the system adjusts its response based on the severity of the threat. This preserves as much of the device's normal operation as possible. Full isolation is reserved for the most high risks. Lower risks receive targeted, less disruptive responses so that a single vulnerable service does not compromise the device's overall functionality. \par

\begin{itemize}
  \item \textit{CRITICAL}: The device is isolated from normal network operation.
  \item \textit{HIGH}: Only the affected service is blocked, leaving the rest of the device reachable.
  \item \textit{MEDIUM}: The affected service is placed under continuous monitoring, maintaining visibility over a latent risk without restricting it.
  \item \textit{LOW}: The finding is logged, and no active enforcement is required.
\end{itemize}

These actions are applied on a per-service basis whenever possible, keeping enforcement as narrowly scoped as the risk allows. However, some vulnerabilities cannot be tied to a specific service. In these cases, the response extends to the entire device. Consequently, a \textit{HIGH} risk may escalate the device to full isolation, effectively matching critical response, since blocking a single service is not possible. Likewise, a non-localized \textit{MEDIUM} finding triggers device-wide monitoring. The risk band still dictates the type of mitigation action; only its scope expands. \par

There are two sources that trigger enforcement, both of which share the same set of actions. The first source is the risk band produced during the assessment process. The second one is runtime monitoring. An alert raised while a device is under continuous monitoring is mapped to the same enforcement actions, allowing the architecture to respond to threats that emerge after a scan. In neither case is enforcement a permanent verdict. Instead, an action remains in place until the next scan reevaluates the associated device. At that point, the action is confirmed, adjusted, or lifted. Together with the scheduled and event-driven reassessments described earlier, this ensures that each device's response remains aligned with its most updated security risk rather than an outdated assessment. \par

Finally, all enforcement is both autonomous and transparent. When the system acts, it generates an incident record that documents the vulnerability and its recommended mitigation. This divides the response into two phases: the system provides immediate, automated containment to limit exposure, and a human operator is responsible for the longer-term task of patching the underlying flaw. Thus, the architecture manages threats without human intervention while keeping human administrators informed and on the complete control of permanent risk mitigation. The next Section debates the deployment of our framework.\par

\section{Implementation}
\label{sec:implementation}

\begin{figure*}[!t]
\centering
\includegraphics[width=\textwidth]{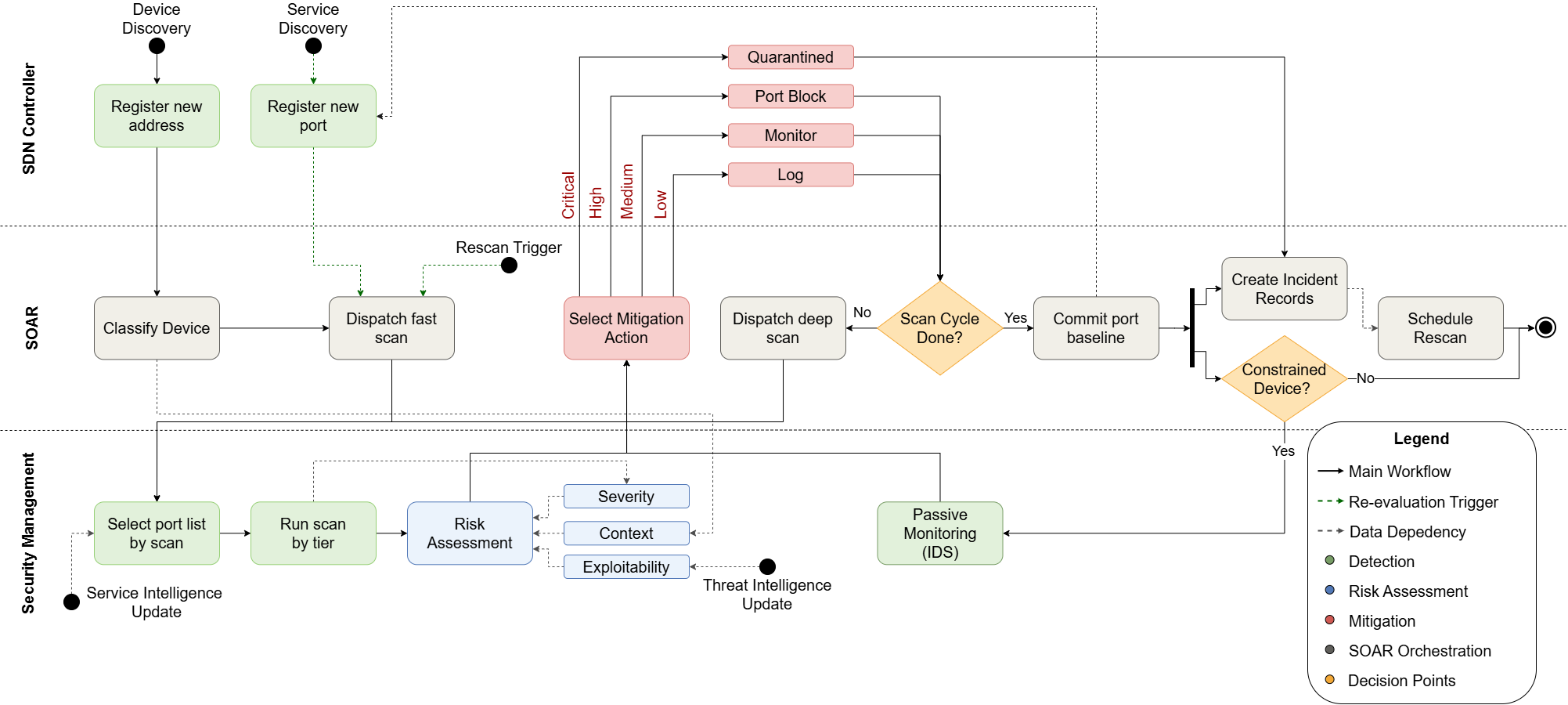}
\caption{\textbf{Deployment of proposed SOAR-driven vulnerability management framework with SDN enforcement.}}
\label{fig:deployment}
\end{figure*}

\noindent This section presents the implementation of the proposed framework, whose deployment architecture is illustrated in Figure~\ref{fig:deployment}. It first describes the deployed environment and the enabling technologies, and then details the three stages of the vulnerability management workflow: vulnerability detection, context-aware risk assessment, and automated mitigation. Finally, it explains how these stages are integrated into a closed-loop orchestration cycle. The implementation focuses on the operational realization of the framework and the interaction among its components, complementing the architectural design presented in Section~\ref{sec:design}. The text below details the purpose of each subsection. 

Subsection~\ref{subsec:impl_testbed} describes the implementation environment, including the software-defined data plane, the orchestrator, the vulnerability scanner, the intrusion detection system, and the network segmentation adopted to isolate different device classes. Subsection~\ref{subsec:impl_detection} presents the vulnerability detection stage, explaining how newly connected devices are classified and how the adaptive two-stage vulnerability assessment strategy is customized according to the device class and current threat intelligence. Subsection~\ref{subsec:impl_assessment} describes the context-aware risk assessment stage, where detected vulnerabilities are assigned a risk band by combining vulnerability severity, exploitation likelihood, and operational context. Finally, Subsection~\ref{subsec:impl_mitigation} presents the automated mitigation stage, which maps each risk band to a proportional SDN-based enforcement action.

\subsection{Proposed Framework}

\label{subsec:impl_testbed}

\noindent The proposed framework was deployed in a programmable network testbed that reproduces a heterogeneous access environment under centralized control. The data plane uses an Open vSwitch (OVS) instance controlled by a Ryu SDN controller, which also serves as the network access control (NAC) layer. Meanwhile, a pfSense gateway provides inter-segment routing, stateful firewall filtering, and DHCP services. This setup enables the SDN controller to detect and classify devices as they join the network. In this way, the controller extracts identifying metadata from the DHCP exchange~\cite{kumar2019all} and forwards it to the orchestrator for device classification. Discovered hosts are classified as either resource-constrained IoT devices (e.g., IoTGoat) or traditional IT hosts (e.g., Metasploitable), enabling tailored scanning, risk assessment, and mitigation strategies. Considering mitigation, as the network is partitioned into isolated VLANs for management, internal/high-exposure endpoints, and quarantine, any incoming classified device is placed in a specific VLAN, reflecting its risk level. For example, high-risk devices are placed or moved to the quarantine VLAN, with full isolation from non high-risk devices.  \par

The Catalyst SOAR platform coordinates the framework, acting as the central orchestrator that guides each device through the vulnerability detection, risk assessment and vulnerability mitigation processes. The technologies behind each stage are listed in Table~\ref{tab:impl_stack} and explained, stage by stage, in the following subsections. \par

\begin{table*}[t]
\centering
\caption{Technologies Used in the Implementation and Their Roles}
\label{tab:impl_stack}
\small
\begin{tabular}{p{3.0cm} p{4.0cm} p{2.0cm} p{7.0cm}}
\hline
\textbf{Component} & \textbf{Technology} & \textbf{Version} & \textbf{Role} \\
\hline
Orchestrator       & Catalyst SOAR   & 0.15.7 & Loop coordination \\
SDN controller     & Ryu             & 4.34 & Access control \& enforcement \\
Data plane         & Open vSwitch    & 3.3.4 & Traffic switching \\
Gateway            & pfSense         & 2.8.1 & Routing, filtering, DHCP \\
Active scanner     & OpenVAS / GVM   & 23.41.6 & Vulnerability scanning \\
Passive IDS        & Suricata        & 8.0.4 & Runtime monitoring \\
Assessment engine  & FastAPI         & 0.118.0 & Risk scoring \& intel fusion \\
Classification     & Fingerbank      & -- & Device-type identification \\
Exploit prediction & EPSS            & 4.0 & Predicted exploit likelihood \\
Confirmed exploits & CISA KEV        & -- & Known-exploited evidence \\
Port relevance     & ISC DShield     & -- & Most-targeted ports \\
Local store        & SQLite          & 3.46.1 & Local caching \\
\hline
\end{tabular}
\end{table*}

\subsection{Detection}
\label{subsec:impl_detection}

\noindent Device classification is the first stage of the system and determines how the subsequent adaptive two-stage vulnerability assessment is performed. In particular, it defines the vulnerability test profile applied to each device, while the scan phase determines the corresponding port set. When a device joins the network, the controller observes its DHCP exchange and extracts three pieces of metadata \cite{kumar2019all}. The first is the Parameter Request List (DHCP Option 55), whose ordering forms a fingerprint of the device's network stack. The second is the Vendor Class Identifier (DHCP Option 60), which carries the vendor's declaration of the device type. The third is the device's MAC address. The controller then forwards these pieces of metadata to the orchestrator, which queries an external fingerprinting service (Fingerbank~\footnote{\url{https://www.fingerbank.org/}}) to resolve them into a device identity. \par

The system matches based on the device's family lineage, the chain of parent categories returned by the service, rather than its specific name, which is too specific to group reliably. A device is labeled \textit{resource-constrained} if any category in its lineage falls within a predefined set of constrained classes. These classes cover families such as IoT, medical, VoIP, printers, and industrial or embedded platforms. Otherwise, it is labeled \textit{traditional IT}. The fingerprinting service only acts as an enabler. It provides the classification label, but the framework treats that label as an input, not a verified result. The classifier was not evaluated here, and other approaches could be used in its place, such as, Link Layer Discovery Protocol (LLDP). LLDP is a vendor-neutral Layer 2 protocol that allows a network asset to advertise its identity and capabilities on the local network, enabling a centralized orchestrator to discover new devices and their characteristics.  \par

After a device is classified, the orchestrator sends a scan request to the security management component (see Figure~\ref{fig:deployment}). This component mediates all communication with the vulnerability scanner (OpenVAS): i) it translates the orchestrator's request into a scan task; ii) it selects the appropriate port list and vulnerability test profile for the device's class; iii) it invokes the necessary scans on OpenVAS; and iv) it returns scan reports to SOAR. Centralizing this scan logic in to the security management component keeps the orchestrator (SOAR) independent of the scanner's interface and concentrates the class-dependent scan configuration on a single place. \par

Two independent choices shape a scan (see Figure~\ref{fig:deployment}, Security Management, "Select port list by scan" and "Run scan by tier"): which ports it covers, set by the scan phase, and which vulnerability tests it runs, set by the device class. The fast scan's port list is not fixed but rebuilt from current attack intelligence, so inspection stays focused on the ports that matter most at any given time. The source of that intelligence is the Internet Storm Center (ISC) DShield project\footnote{\url{https://www.dshield.org/}}, a public feed that aggregates firewall and intrusion logs from sensors worldwide and reports, per day, the ports receiving the most attack activity. From this feed the system selects the most-targeted ports over a sliding window of recent activity. A single window, however, biases the result: a short one captures ports spiking recently but misses those attacked persistently over in the past; while a long window does the reverse. To avoid favouring either, the system unions three tiers, weekly, monthly, and yearly, so the port list reflects both newly emerging and persistently targeted ports. The full port list construction, including the deep list, is summarized in Algorithm~\ref{alg:port_selection}. \par

Each selected port is then classified by transport. Because a port can be attacked over TCP, UDP, or both, the system keeps a port on a transport only when that transport carries a significant share of its attack activity, which in our deployment is 30\%. The share is derived from the historical record of attacks observed on the port, so the split reflects how the port is actually targeted rather than an assumption about its default protocol. The result is a compact, transport-aware port list that keeps the fast scan both relevant and agile. \par

\begin{algorithm}[htbp]
\caption{Threat-informed port list construction}
\label{alg:port_selection}
\begin{algorithmic}[1]
\REQUIRE Per-day most-attacked port records from the public feed; the three window lengths (weekly, monthly, yearly); the per-tier size $N$; the transport-share threshold
\ENSURE A fast port list and a complementary deep port list, each transport-aware

\STATE \textit{// Fast list: union of three time tiers}
\FOR{each tier in \{weekly, monthly, yearly\}}
    \STATE rank ports by attack volume within that tier's window
    \STATE take the top $N$ ports of the tier
\ENDFOR
\STATE $\text{fast} \gets$ union of the three tiers' top-$N$ ports

\STATE \textit{// Transport classification}
\FOR{each port in $\text{fast}$}
    \STATE keep it on TCP if TCP is a significant share of its activity
    \STATE keep it on UDP if UDP is a significant share of its activity
\ENDFOR

\STATE \textit{// Deep list: complement of the fast list}
\STATE $\text{deep} \gets$ (all TCP ports $\setminus$ fast TCP) $\cup$ (top attack-ranked UDP ports $\setminus$ fast UDP)

\RETURN $\text{fast}$, $\text{deep}$
\end{algorithmic}
\end{algorithm}

The choice of vulnerability tests is the second axis, and it is tailored to the device class. Since devices are discovered as they join the network without prior enrollment, all scanning is unauthenticated, leaving no credentials with which to authenticate to the targets. Traditional IT devices receive the full test set, which includes active vulnerability tests that probe services directly to confirm flaws. Resource-constrained devices receive only the reconnaissance group. These tests infer device vulnerabilities from the information the scanned services disclose: once a service and its version are identified, the scanner matches them against its vulnerability data to flag known issues. Every constrained device is therefore assessed on what it reveals about itself rather than on how it responds to an attack, which keeps the scan safe to run from start to finish on fragile endpoints, although it yields more conservative results because it only detects flaws through the service version. \par

On top of this class split, the bridge applies a second, finer selection to the individual tests themselves. Each Network Vulnerability Test (NVT) targets a known vulnerability, and the bridge admits a test only when the risk band of its target vulnerability meets a configurable threshold, pruning the rest before the scan runs (see Figure~\ref{fig:deployment}, Security Management, "Select port list by scan"). This NVT filtering narrows the plugin set to the vulnerabilities that matter most under the current risk policy, so scan time is not spent confirming flaws the policy would not act on. The threshold is adjustable rather than static, and its effect is examined in Section~\ref{sec:evaluation}. \par

The two scan phases run sequentially and are designed so that the deep scan picks up where the fast scan leaves off. First, the fast scan runs over the focused port list, allowing the most likely exposures to surface quickly. If its findings do not warrant stopping the process, a deep scan follows to widen coverage over the complementary port range defined in Algorithm~\ref{alg:port_selection}. Because the two lists do not overlap, the deep scan examines only what the fast scan did not, and together the two phases cover the device's relevant attack surface. \par

For resource-constrained devices, the detection process does not end with the scan. Once the two-phase scan cycle is complete, Suricata, an intrusion detection system (IDS) \cite{diana2025overview}, places the network devices under continuous passive monitoring. Suricata inspects their traffic without injecting anything of its own. This monitoring also covers services that the system intentionally keeps reachable when a finding is serious enough to monitor but not serious enough to block. The aim in both cases is to keep a service in use while retaining visibility over it. For constrained devices, this makes up for the limited information that a reconnaissance-only scan provides. For a service that is kept in use, monitoring acts as a safeguard so that, if a threat begins to materialize, the IDS can detect it. The system receives severity-ranked alerts from Suricata and is notified of threats that emerge after the scan. After that, if necessary, the system can perform mitigation actions. \par

Passive monitoring is not the only method used to watch a device after it is scanned. The controller also tracks the set of ports that each device presented during its last scan. This set is recorded as a baseline when the scan cycle is complete. When a device later sends traffic from a port outside of this baseline, the controller considers it a change in the device's exposure and triggers a quick scan on the Security Management (see Figure~\ref{fig:deployment}, SDN Controller, "Register new port"). The newly observed port is added to the port list of the scan, ensuring that it is always inspected, even if it does not appear in the ISC top-ranked ports list. This transforms a change on device's behavior into a detection trigger, enabling the identification and assessment of services that become active after onboarding rather than waiting for the next scheduled scan by the orchestrator. \par

\subsection{Risk Assessment}
\label{subsec:impl_assessment}

\noindent The risk assessment runs on the report returned by the scanner in the security management, so the orchestrator receives findings that are already banded and never computes risk itself. The scanning process uses the Greenbone Vulnerability Manager (GVM)\footnote{\url{https://www.greenbone.net/}} with the OpenVAS scanner, and each finding carries a technical severity score based on the Common Vulnerability Scoring System (CVSS) \cite{mell2006common}. The security management scores each finding on two axes: severity, based on the CVSS score, and exploitability, based on two threat intelligence sources. These are the Exploit Prediction Scoring System (EPSS) \cite{jacobs21}, a predicted probability of exploitation, and the CISA Known Exploited Vulnerabilities (KEV) catalog of vulnerabilities confirmed as exploited in the wild \footnote{https://www.cisa.gov/known-exploited-vulnerabilities-catalog}. Both sources of intelligence are held in a local SQLite database that is refreshed daily, so the scoring reflects the intelligence available at the time of the last refresh rather than a live query for each finding.

Severity is categorized into three tiers (high, moderate, and low) based on the CVSS breakpoints in Table~\ref{tab:severity_exploitability}. In turn, exploitability is resolved into a two-level axis, high or low. It is considered high if any of a finding's CVEs are listed in KEV or if the highest EPSS probability among its CVEs is at or above the cutoff. A finding is considered to have low exploitability when its CVEs are present in the EPSS feed and score below the cutoff without a KEV listing. If there is no exploitability evidence, such as if a finding has no CVEs or if its CVEs appear in neither the KEV nor the EPSS feed, then it is treated as high by default. The EPSS cutoff is a configurable parameter that matches the organization's risk tolerance. A lower cutoff raises more findings to "high", while a higher cutoff reserves it for the most probable threats. 

\begin{table}[htbp]
\caption{Severity and Exploitability Matrix}
\label{tab:severity_exploitability}
\centering
\renewcommand{\arraystretch}{1.3} 
\begin{tabular}{lcc}
\toprule
\multirow{2}{*}{\textbf{Severity (CVSS)}} & \multicolumn{2}{c}{\textbf{Exploitability (EPSS / KEV)}} \\
\cmidrule(lr){2-3}
& \textbf{High} & \textbf{Low} \\
\midrule
CVSS $\geq 7.0$               & Critical & Medium \\
$4.0 \leq \text{CVSS} < 7.0$  & High     & Low \\
CVSS $< 4.0$                  & Low      & Low \\
\bottomrule
\end{tabular}
\end{table}

The severity tier and exploitability level are combined using the matrix in Table~\ref{tab:severity_exploitability} to assign each finding to one of four risk bands. Algorithm~\ref{alg:band_assignment} summarizes the underlying exploitability logic. The matrix is asymmetric to ensure that confirmed or highly probable exploitation outweighs raw technical severity. For instance, an exploited moderate-severity finding is assigned a higher risk band than an unexploited high-severity finding. Furthermore, risk is assessed at the finding level rather than being aggregated into a single, host-wide verdict. Since the Security Management outputs a list of individually banded findings, every affected service retains its own distinct risk band. This enables subsequent enforcement actions to target specific services independently.

\begin{algorithm}[htbp]
\caption{Context-aware band assignment for a finding}
\label{alg:band_assignment}
\begin{algorithmic}[1]
\REQUIRE Finding $v$ with a CVSS severity; the CVEs of $v$; the EPSS feed; the CISA KEV catalog; the host context attributes
\ENSURE $v$ is assigned a final risk band

\STATE \textit{// Exploitability: high or low}
\IF{$v$ has at least one CVE scored in the EPSS feed}
    \STATE $\text{kev\_listed} \gets (\text{some CVE of } v \text{ is in KEV})$
    \STATE $\text{epss\_high} \gets (\text{max EPSS} \geq \text{cutoff})$
    \STATE $\text{exploit\_high} \gets \text{kev\_listed} \lor \text{epss\_high}$
\ELSE
    \STATE $\text{exploit\_high} \gets \mathrm{true}$ \COMMENT{no evidence: worst-case default}
\ENDIF

\STATE \textit{// Technical band from the severity--exploitability matrix (Table~\ref{tab:severity_exploitability})}
\STATE $\text{band} \gets \text{matrix}(\text{severity}(v),\ \text{exploit\_high})$

\STATE \textit{// Context promotion: single bounded step}
\IF{host of $v$ meets any context condition (Table~\ref{tab:context_promotion})}
    \STATE $\text{band} \gets$ one band higher, capped at the top band
\ENDIF

\RETURN $\text{band}$
\end{algorithmic}
\end{algorithm}

The band is then adjusted using configurable context (Figure~\ref{fig:deployment}, Security Management, "Risk Assessment"), allowing operators to promote findings based on host exposure or criticality. In our deployment, findings are promoted by one band if the host is in a sensitive network zone or is a constrained device (Table~\ref{tab:context_promotion}). Promotion is applied once, regardless of how many conditions are met, and capped at the highest band. The resulting bands are forwarded to the SOAR orchestrator, which maps each band to an enforcement action, as described in the next subsection.

\begin{table}[htbp]
\caption{Context-Based Band Promotion}
\label{tab:context_promotion}
\centering
\renewcommand{\arraystretch}{1.3}
\begin{tabular}{lc}
\hline
\textbf{Context Condition} & \textbf{Promotion} \\
\hline
Sensitive zone      & +1 \\
Normal zone         & 0 \\
Constrained device  & +1 \\
Traditional device  & 0 \\
\hline
\end{tabular}
\end{table}

\subsection{Mitigation}
\label{subsec:impl_mitigation}

\noindent Mitigation enforces the severity level assigned to each finding during the assessment stage. It translates a band into a proportional action on the data plane (see Figure~\ref{fig:deployment}, SOAR, "Select Mitigation Action"). There are four bands and four corresponding actions. A critical finding quarantines the host. A high finding blocks the affected service. A medium finding places the service under monitoring. A low finding is logged with no active enforcement. \par

Two rules govern how these actions combine on a host, as set out in Algorithm ~\ref{alg:mitigation}. First, quarantine is exclusive: a critical finding isolates the entire device, so any other planned actions for that host are canceled, since a quarantined device is unreachable.The second rule is that enforcement is per service otherwise. Since each finding retains its own band, a single host can have several actions at once. For example, there could be an L4 block on one port and monitoring on another, each scoped to the service it concerns. This keeps the response as narrow as the evidence allows so that one vulnerable service does not affect the device's other functions. When a finding cannot be tied to a specific service, the scope of the action widens to the entire host: a high finding with no identifiable port escalates to quarantine, and a medium finding with no identifiable port becomes full-host monitoring. Therefore, the severity level determines the type of action and whether the finding can be localized determines its scope. \par

\begin{algorithm}[htbp]
\caption{Band-to-action enforcement for a host}
\label{alg:mitigation}
\begin{algorithmic}[1]
\REQUIRE The findings of a host, each with a risk band and an optional service port
\ENSURE The set of enforcement actions applied to the host
\IF{any finding is CRITICAL, or is HIGH with no identifiable port}
    \STATE quarantine the host \COMMENT{exclusive: clears other actions}
    \RETURN
\ENDIF
\STATE actions $\gets \emptyset$
\FORALL{findings $v$ of the host}
    \IF{band$(v)$ is HIGH}
        \STATE add an L4 block on the port of $v$
    \ELSIF{band$(v)$ is MEDIUM}
        \STATE add monitoring on the port of $v$, or full-host monitoring if $v$ has no port
    \ELSE
        \STATE logged $v$
    \ENDIF
\ENDFOR
\RETURN actions
\end{algorithmic}
\end{algorithm}

Each action is enforced through the SDN controller. Quarantine reassigns the host to a dedicated quarantine segment, which takes effect when the host renews its lease. Its existing flows are terminated so the isolation applies immediately. An L4 block installs a targeted drop rule for a specific protocol and port, leaving the rest of the device reachable. Monitoring installs a SPAN mirror that copies relevant traffic to Suricata for passive inspection without restricting it. A logged finding produces no data-plane action. Thus, the enforcement granularity matches the risk: full isolation for the most severe case and a blocked or mirrored service for the rest. \par

There are two sources of enforcement that share this same vocabulary. The first source is the band produced by a scan, as described above. The second is the runtime IDS, which maps an alert raised by Suricata while a device is under monitoring onto the same bands and drives the same actions. This allows the system to respond not only to what a scan detects, but also to threats that emerge after the scan is complete. \par

No action is permanent. Each one remains in effect until the next scan re-evaluates the host or a new service trigger arises. At that time, it is confirmed, adjusted, or lifted according to the device's current state. The reevaluations are scheduled per device, and the interval is a configurable policy. In our deployment, a traditional device is rescanned every 50 hours, while a constrained device is rescanned every 100 hours. Constrained devices are rescanned less often because they are under continuous passive monitoring between scans. This monitoring provides visibility without the load of repeated active probing. Traditional devices, on the other hand, rely on more frequent active rescans. \par

Finally, the findings are recorded for the human operator (see Figure~\ref{fig:deployment}, SOAR, "Create Incident Records"). Each finding in the GVM report has mitigation text supplied by the scanner. The orchestrator's ticket handler organizes this output into two records. The first record documents the findings and the band to which each vulnerability was assigned. This provides a per-service view of the host's risk. The second record is a mitigation support record that collects the mitigation text from the scanner for the affected vulnerabilities. This provides the operator with concrete guidance for the follow-up task of patching after the automated containment is already in place. \par

The next section discusses the evaluation results obtained from a set of experiments run on the proposed vulnerability management framework.  \par

\section{Evaluation}
\label{sec:evaluation}

This section evaluates the proposed SOAR-driven vulnerability management framework with respect to the research questions introduced in Section~\ref{sec:introduction}. The objective is to assess whether the proposed approach improves the efficiency and effectiveness of vulnerability management by combining adaptive vulnerability discovery, context-aware risk assessment, and automated mitigation through SDN.

The evaluation is organized into several complementary tests. Section~\ref{sec:experimental_setup} describes the experimental environment and evaluation methodology adopted throughout all the tests. Sections~\ref{sec:adaptive_scan_cost} and~\ref{sec:constrained_aware_scanning} evaluate the two-stage scanning strategy, quantifying its scanning overhead and its effectiveness in balancing vulnerability coverage against the constraints of heterogeneous devices. Section~\ref{sec:pipeline_latency} measures the end-to-end processing latency of the proposed framework, assessing its suitability for near real-time automatic vulnerability management. Section~\ref{sec:risk_band_assignment} investigates the behavior of the contextual risk assessment model by analyzing how CVSS severity, exploitation likelihood, and operational context influence the assigned risk bands. Building upon this qualitative analysis, Section~\ref{sec:exploitation_aware_risk} statistically validates the proposed context-aware prioritization strategy, demonstrating its ability to reduce unnecessary mitigation actions while preserving complete coverage of vulnerabilities with verified exploitation. Finally, Section~\ref{sec:scalability} analyzes the scalability of the proposed framework under increasing numbers of managed assets and concurrent security management events.

Collectively, these six experiments provide comprehensive qualitative and quantitative evidences that the proposed framework improves vulnerability management by enabling adaptive scan, accurate risk prioritization, efficient automated mitigation, and scalable operation across heterogeneous network environments, while significantly reducing the workload of security teams.

\subsection{Experimental Setup}
\label{sec:experimental_setup}

This section presents the experimental evaluation conducted to answer the four research questions (RQs) introduced in Section~\ref{sec:introduction}. Since the proposed framework integrates adaptive vulnerability discovery, context-aware risk assessment, automated risk band-based mitigation, and scalable vulnerability management orchestration, the evaluation is organized into a set of complementary experiments, each assessing a relevant specific aspect of the framework. Collectively, these experiments provide qualitative and quantitative evidence of the framework's effectiveness, efficiency, and scalability. 

Tables~\ref{tab:rq-coverage} and~\ref{tab:testbed-coverage} provide an overview of the framework evaluation. The first table relates each experiment to the research question(s) it addresses, whereas the second table identifies the experimental testbed used to conduct each evaluation. As shown in Table~\ref{tab:testbed-coverage} two testbed types have been used: virtualized (A) and hybrid (B). The last testbed type has joined physical and virtualized assets. The virtualized testbed (A) is the fully virtualized software-defined environment described in Section~\ref{sec:implementation}, with no physical hosts. The hybrid testbed (B) extends it with physical laboratory machines that host the target virtual machines, and is used for the concurrent-load scalability experiments. Table~\ref{tab:target_vms} summarizes the technical specifications of the three target virtual machines, while Table~\ref{tab:server_spec} presents the specifications of the scanner host.

\begin{table}[h!]
\centering
\caption{Mapping of Evaluation Tests to Research Questions.}
\label{tab:rq-coverage}
\footnotesize
\setlength{\tabcolsep}{3pt}
\renewcommand{\arraystretch}{1.15}
\begin{tabular}{p{4.2cm}cccc}
\hline
\textbf{Evaluation Test} & \textbf{RQ1} & \textbf{RQ2} & \textbf{RQ3} & \textbf{RQ4} \\
\hline
Adaptive Scan Cost \& Coverage Trade-off     & x       & --      & --      & -- \\
Constrained-Aware Scanning \& Safety-Depth   & x       & --      & --      & -- \\
Pipeline Latency                             & --      & --      & x       & x  \\
Risk Band Assignment                         & --      & x       & --      & -- \\
Exploitation-Aware Risk Prioritization       & --      & x       & --      & -- \\
Scalability                                  & --      & --      & --      & x  \\
\hline
\end{tabular}
\end{table}

\begin{table}[h!]
\centering
\caption{Mapping of Evaluation Tests to Testbed Types.}
\label{tab:testbed-coverage}
\footnotesize
\setlength{\tabcolsep}{3pt}
\renewcommand{\arraystretch}{1.15}
\begin{tabular}{p{4.2cm}cc}
\hline
\textbf{Evaluation Test} & \textbf{Virtualized (A)} & \textbf{Hybrid (B)} \\
\hline
Adaptive Scan Cost \& Coverage Trade-off     & x      & -- \\
Constrained-Aware Scanning \& Safety-Depth   & x      & -- \\
Pipeline Latency                             & x      & -- \\
Risk Band Assignment                         & x      & -- \\
Exploitation-Aware Risk Prioritization       & x      & -- \\
Scalability                                  & --     & x  \\
\hline
\end{tabular}
\end{table}

\begin{table}[htbp]
\centering
\caption{Hardware and Software Specifications of the Target Virtual Machines.}
\label{tab:target_vms}
\footnotesize
\renewcommand{\arraystretch}{1.3}
\setlength{\tabcolsep}{4pt}
\begin{tabular}{lccc}
\hline
\textbf{Specification} & \textbf{Metasploitable2} & \textbf{IoTGoat} & \textbf{Windows 2003} \\
\hline
OS   & Ubuntu 8.04 & OpenWrt & Windows 2003 \\
CPU  & 1           & 1       & 1            \\
RAM  & 512 MB      & 256 MB  & 1 GB         \\
\hline
\end{tabular}
\end{table}

\begin{table}[htbp]
\centering
\caption{Hardware and Software Specifications of the Scanner Host.}
\label{tab:server_spec}
\footnotesize
\renewcommand{\arraystretch}{1.3}
\setlength{\tabcolsep}{4pt}
\begin{tabular}{lcc}
\hline
\textbf{Specification} & \textbf{Host Machine} & \textbf{Scanner VM} \\
\hline
OS   & Windows 11 Pro  & Kali Linux [2026.1] \\
CPU  & Intel i9-14900HX & 8 CPU Cores \\
RAM  & 32 GB      & 24 GB \\
\hline
\end{tabular}
\end{table}

\subsection{Adaptive Scan Cost and Coverage Trade-off}
\label{sec:adaptive_scan_cost}

\noindent This test addresses RQ1, which is related to the experimental assessment in how the scan should be configured. The test evaluates the two configurable filters that shape a scan, the risk threshold applied to the vulnerability tests, and the list of ports covered by the scan. For each factor, the test asks the same question: Does narrowing the scan reduce its cost enough to justify the coverage it gives up? The reported results were obtained from a Metasploitable2 host and represent the average of three test runs. \par

The first configuration filter determines which NVTs run by admitting only those whose target vulnerability meets a minimum risk threshold. This filter only applies to vulnerability tests that can be ranked by risk. These tests are matched through KEV, EPSS, or are left as no-CVE. Diagnostic helper tests carry no risk band and are always retained, so they are excluded from the reduction. Increasing the threshold eliminates a portion of the filterable set. Compared to the unfiltered configuration, the number of ranked tests decreases by approximately 23\% at the MEDIUM threshold and by up to 60\% at the CRITICAL threshold, as shown in Table~\ref{tab:nvt_filter}. However, this reduction does not translate into a meaningful improvement in scan duration, which remains practically unchanged across all thresholds. Since the filter narrows the test set without shortening the scan, there is no cost savings to balance against the coverage that a higher threshold would eliminate. For this reason, the framework operates at the lowest threshold, which admits the full test set and preserves the complete baseline. The risk filter remains available but is left unused since narrowing the test set here provides no evident benefit. \par

\begin{table}[t]
\centering
\caption{NVT Selection Under Increasing Risk Threshold}
\label{tab:nvt_filter}
\footnotesize
\setlength{\tabcolsep}{3pt}
\renewcommand{\arraystretch}{1.15}
\resizebox{\columnwidth}{!}{
\begin{tabular}{p{1.4cm}rrrrrrc}
\hline
\textbf{Config.} & \textbf{Help.} & \textbf{KEV} & \textbf{EPSS} & \textbf{No-CVE} & \textbf{Total} & \textbf{Red.\%} & \textbf{Time (m)} \\
\hline
LOW      & 16559 & 107 & 2109 & 1465 & 20240 & 0.0  & 43 \\
MEDIUM   & 16559 & 107 & 1344 & 1385 & 19395 & 23.0 & 42 \\
HIGH     & 16559 & 107 & 779  & 1385 & 18830 & 38.3 & 45 \\
CRITICAL & 16559 & 102 & 561  & 811  & 18033 & 60.0 & 41 \\
\hline
\end{tabular}}
\end{table}

The second configuration filter determines the port list, where the trade-off is reversed. Instead of scanning the entire port range, the fast scan focuses only on the ports most frequently targeted by attackers, as identified by the ISC DShield feed. Table~\ref{tab:port_reduction} summarizes the impact of different port-list sizes. In contrast to the NVT filter, reducing the number of scanned ports yields a substantial reduction in scan time. The Top 200 configuration scans only 295 ports instead of the full port range, reducing the scan duration by approximately 50\% while still discovering 18 of the 30 open ports identified by the baseline scan. As the list expands to the Top 500 and Top 1,000 ports, additional open ports are detected (22 and 25, respectively), but the time savings decrease considerably, from approximately 50\% for the Top 200 configuration to about 13\% for the Top 1,000 configuration, as the scan duration gradually approaches that of the baseline.

\begin{table}[t]
\centering
\caption{Port List Reduction via ISC DShield}
\label{tab:port_reduction}
\footnotesize
\setlength{\tabcolsep}{3pt}
\renewcommand{\arraystretch}{1.15}
\resizebox{\columnwidth}{!}{
\begin{tabular}{p{3.2cm}rrrrc}
\hline
\textbf{Configuration} & \textbf{TCP} & \textbf{UDP} & \textbf{Total} & \textbf{Open Ports} & \textbf{Time (m)} \\
\hline
All TCP + Top 100 UDP    & 65535  & 100  & 65635 & 30 & 45 \\
ISC DShield - Top 1000   & 1281   & 137  & 1418  & 25 & 39 \\
ISC DShield - Top 500    & 618    & 94   & 712   & 22 & 32 \\
ISC DShield - Top 200    & 239    & 56   & 295   & 18 & 22 \\
\hline
\end{tabular}}
\end{table}

A short list remains effective because open-port coverage and vulnerability coverage are not the same, and vulnerability coverage is what ultimately determines the quality of the assessment. Table~\ref{tab:port_coverage} shows the percentage of the baseline's results that each list includes, grouped by risk band. Although the Top 200 risk band discovers only a subset of the open ports identified by the baseline scan, it still retains approximately 91\% of all detected vulnerabilities, including nearly 90\% of critical vulnerabilities and more than 91\% of high-risk findings. As the port list expands to the Top 500 and Top 1,000 ports, vulnerability coverage increases progressively, reaching almost complete coverage with the Top 1,000 configuration. This behaviour indicates that vulnerabilities are concentrated on the ports most frequently targeted by attackers, which are precisely the ports prioritized by the ISC DShield ranking. Consequently, the ports omitted by shorter lists contribute relatively little to the overall vulnerability coverage. Therefore, the Top 200 configuration provides the best trade-off between assessment time and vulnerability coverage, preserving most of the relevant findings while achieving the largest reduction in scan duration.

\begin{table}[t]
\centering
\caption{Vulnerability Coverage by Risk Band}
\label{tab:port_coverage}
\footnotesize
\setlength{\tabcolsep}{3pt}
\renewcommand{\arraystretch}{1.15}
\resizebox{\columnwidth}{!}{
\begin{tabular}{p{3.2cm}rrrrr}
\hline
\textbf{Configuration} & \textbf{Crit.} & \textbf{High} & \textbf{Med.} & \textbf{Low} & \textbf{Final} \\
\hline
All TCP + Top 100 UDP    & 100\%   & 100\%   & 100\%   & 100\%  & 100\%  \\
ISC DShield - Top 1000   & 98.0\%  & 100\%   & 100\%   & 100\%  & 99.8\% \\
ISC DShield - Top 500    & 93.9\%  & 96.6\%  & 94.7\%  & 89.3\% & 94.8\% \\
ISC DShield - Top 200    & 89.8\%  & 91.3\%  & 91.8\%  & 89.3\% & 91.1\% \\
\hline
\end{tabular}}
\end{table}

Considering the results discussed along this subsection, the framework adopts the Top 200 port list for the fast scan. This choice is driven by the objective of minimizing the time to mitigation. By completing in approximately half the time required for a full-range scan, the fast stage enables the framework to identify the initial security posture of newly admitted devices and apply proportional mitigation actions significantly earlier. This proactive response reduces the window of exposure while the device remains insufficiently assessed. The reduced port coverage of the fast scan does not compromise the overall risk assessment, as the subsequent deep scan examines the remaining port range and recovers the small vulnerability gap introduced by the Top 200 configuration.

\subsection{Constrained-Aware Scanning and the Safety-Depth}
\label{sec:constrained_aware_scanning}

\noindent This test addresses RQ1 by comparing the constrained-aware profile, which performs only reconnaissance tests, with the full profile used for traditional IT devices. Measurements for the constrained-resource device case were collected from an IoTGoat host and averaged over three runs. Table~\ref{tab:constrained_scan} summarizes the results.

\begin{table*}[t]
\centering
\caption{Effect of Scan Profile and Port List on a Constrained Device}
\label{tab:constrained_scan}
\footnotesize
\renewcommand{\arraystretch}{1.2}
\setlength{\tabcolsep}{5pt}
\begin{tabular}{llrrrrrrrrr}
\hline
\textbf{NVT Profile} & \textbf{Scan Profile} & \textbf{TCP} & \textbf{UDP} & \textbf{Total Ports} & \textbf{Scan Time (min)} & \textbf{Crit} & \textbf{High} & \textbf{Med} & \textbf{Low} & \textbf{Coverage} \\
\hline
Full  & Full range        & 65535 & 100 & 65635 & 151 & 1 & 2 & 2 & 2 & 100\%  \\
Aware & Full range        & 65535 & 100 & 65635 & 60  & 0 & 2 & 1 & 2 & 71.4\% \\
\arrayrulecolor{gray!20}\hline
Aware & Fast (Top 200)    & 239   & 56  & 295   & 14  & 0 & 2 & 1 & 2 & 71.4\% \\
Aware & Deep (complement) & 65296 & 44  & 65340 & 47  & 0 & 0 & 1 & 2 & 42.9\% \\
\hline
Full  & Fast (Top 200)    & 239   & 56  & 295   & 62  & 0 & 2 & 1 & 2 & 71.4\% \\
\arrayrulecolor{black}
\hline
\end{tabular}
\end{table*}

Over the same ports, the aware profile completes in 60 minutes against 151 for the full profile. It retains 71.4\% of the vulnerabilities identified by the full profile. The difference comes from the tests each profile runs. The full profile includes active tests that probe a service directly to confirm a vulnerability, whereas the aware profile excludes these tests and only includes reconnaissance tests that read what a service reveals about itself. This is what makes it safe on a fragile device, but it also means the aware profile misses anything that only an active test would confirm. On this host that missing vulnerability coverage is one critical and one medium finding, and the critical is a backdoor (ingreslock) on 5515/tcp. This service does not expose a version banner and can only be confirmed by active exploitation, so a reconnaissance-only scan cannot detect it. This limitation is inherent to reconnaissance-only scanning and is explicitly addressed in the proposed framework, as following explained. Following the active scan, constrained devices remain under continuous passive monitoring, allowing the framework to detect suspicious behaviour associated with threats that cannot be confirmed through safe reconnaissance alone. \par

Restricting the aware profile to the top 200 ports reduces the scan time to 14 minutes while maintaining the same 71.4\% vulnerability coverage as the full-range aware scan. Although the device is not fully assessed in 14 minutes, the fast scan has already found enough for the system to establish an initial security posture and apply timely and proportional mitigation actions. Although both configurations inspect the same port set, the full profile requires 62 minutes compared with only 14 minutes for the aware profile because of the additional active verification tests. Running the full profile over the same top 200 ports achieves the same 71.4\% coverage. This relevant observation demonstrates that the reduced vulnerability coverage originates from the use of the reconnaissance-only NVT profile rather than from limiting the scan to the Top 200 ports.

The deep scan, which uses the aware profile, covers the remaining ports in 47 minutes. Its reported vulnerabilities partially overlap those of the fast scan because both stages identify host-level findings that are independent of the scanned port set. The total time is what it matters here. Together, the fast and deep scans take about 61 minutes, practically the same as the 60 minutes it takes for a single Aware scan of the full range. Therefore, splitting the scan into two stages costs almost no extra time yet lets the system act after 14 minutes instead of a full hour, reducing the time required to establish an actionable security posture by 77\%.\par

Taken together, these results illustrate the most important trade-offs for constrained devices. The aware profile maintains normal device operation and preserves most of the vulnerability coverage in a fraction of the time. The two-stage scan split provides an early response without extending the overall scan time. The posterior continuous passive monitoring compensates for this limitation, making the proposed strategy preferable for fragile devices, where an exhaustive active scan would require more than two hours and could disrupt their normal operation. \par

\subsection{Pipeline Latency}
\label{sec:pipeline_latency}

\noindent This test addresses RQ3 and RQ4. It evaluates the operational efficiency of the proposed framework by quantifying the latency and computational cost associated with its main processing stages. Specifically, the evaluation comprises three complementary analyses. First, it measures the end-to-end assessment latency, from host discovery to the completion of the vulnerability and risk assessment pipeline (Table~\ref{tab:pipeline_latency}). Second, it quantifies the recurring background overhead associated with maintaining the threat-intelligence and port-relevance local repositories, which is incurred independently of any vulnerability scan (Table~\ref{tab:periodic_cost}). Finally, it evaluates the enforcement latency, defined as the elapsed time between a mitigation decision and the corresponding rule(s) becoming active on the data plane (Table~\ref{tab:enforcement_latency}). Unless otherwise stated, each reported value corresponds to the mean of 20 independent experimental runs.

\begin{table}[htbp]
\caption{Per-Stage Pipeline Latency}
\label{tab:pipeline_latency}
\centering
\footnotesize
\setlength{\tabcolsep}{3pt}
\renewcommand{\arraystretch}{1.2}
\begin{tabular}{p{0.6cm}p{3.2cm}c}
\hline
\textbf{\#} & \textbf{Stage} & \textbf{Mean (s)} \\
\hline
1 & Detection \& Classification & 6.52 \\
2 & Dispatch                    & 3.15 \\
3 & Scan Preparation            & 1.42 \\
4 & \textit{Scan Execution}     & \textit{see Tables~\ref{tab:port_reduction} and \ref{tab:constrained_scan}} \\
5 & Risk Assessment             & 5.47 \\
6 & \textit{Mitigation}         & \textit{see Table~\ref{tab:enforcement_latency}} \\
\hline
\end{tabular}
\end{table}

Table~\ref{tab:pipeline_latency} reports the per-stage latency of the critical path. \emph{Detection \& Classification} observes the host's arrival, triggers the reaction, and classifies the device, and \emph{Dispatch} hands the assessment to the downstream components. \emph{Scan Preparation} creates the scan task, tailoring the port list to the scan depth and the vulnerability tests to the device class. \emph{Risk Assessment} retrieves the results, computes the risk band, and derives the proportional response. Scan execution and the mitigation itself are reported separately in Tables~\ref{tab:port_reduction}, \ref{tab:constrained_scan}, and~\ref{tab:enforcement_latency}.

Excluding scan execution, most measured framework overhead arises from coordinating sequential SOAR actions across components. The platform executes each reaction as a separate script, introducing inter-component coordination delays. Consequently, the current orchestration model is the principal source of non-scanning pipeline latency and its main operational limitation.

A host that is not quarantined after the fast scan is escalated to a deep scan, and the deep stage results re-enter the pipeline. In this second pass, \emph{Detection \& Classification} does not run again, since the host is already known and classified, whereas \emph{Dispatch}, \emph{Scan Preparation}, and \emph{Risk Assessment} are repeated. Aggregating the per-stage latencies in Table~\ref{tab:pipeline_latency}, a fast-only assessment introduces an orchestration overhead of 16.56~s (stages~1--3 and~5), which is is approximately 2.0\% of the 14-minute fast scan. A fast-to-deep assessment repeats stages~2, 3, and~5 for the deep pass, adding 10.04~s for a total of 26.60~s, which is approximately 0.73\% of the dual-stage scanning duration. In both cases, the orchestration overhead is always a small fraction of the scan-execution time, which spans tens of minutes (Tables~\ref{tab:port_reduction} and~\ref{tab:constrained_scan}), confirming that the framework's own processing overhead is not the bottleneck on the pipeline latency.

\begin{table}[htbp]
\caption{Periodic Maintenance Cost}
\label{tab:periodic_cost}
\centering
\setlength{\tabcolsep}{3pt}
\renewcommand{\arraystretch}{1.2}
\begin{tabular}{p{0.6cm}p{3.2cm}cc}
\hline
\textbf{\#} & \textbf{Operation} & \textbf{Cold Start (s)} & \textbf{Daily (s)} \\
\hline
1 & NVT Catalogue Sync          & 346.88 & --- \\
2 & EPSS Sync                   & 1.06   & 1.06 \\
3 & KEV Sync                    & 0.72   & 0.72 \\
4 & Attack Telemetry Ingestion  & 152.95 & 8.32 \\
5 & Fast Port List Build        & 24.39  & 24.39 \\
6 & Deep Port List Build        & 59.79  & 59.79 \\
\hline
\multicolumn{2}{l}{\textbf{Total}} & \textbf{585.79} & \textbf{94.28} \\
\hline
\end{tabular}
\end{table}

Table~\ref{tab:pipeline_latency} reports the per-stage latency of the critical path. \emph{Detection \& Classification} observes the host's arrival, triggers the reaction, and classifies the device, and \emph{Dispatch} hands the assessment to the downstream components. \emph{Scan Preparation} creates the scan task, tailoring the port list to the scan depth and the vulnerability tests to the device class. \emph{Risk Assessment} retrieves the results, computes the risk band, and derives the proportional response. Scan execution and the mitigation itself are reported separately in Tables~\ref{tab:port_reduction}, \ref{tab:constrained_scan}, and~\ref{tab:enforcement_latency}.

Excluding scan execution, most measured framework overhead arises from coordinating sequential SOAR actions across components. The platform executes each reaction as a separate script, introducing inter-component coordination delays. Consequently, the current orchestration model is the principal source of non-scanning pipeline latency and its main operational limitation.

A host that is not quarantined after the fast scan is escalated to a deep scan, and the deep stage results re-enter the pipeline. In this second pass, \emph{Detection \& Classification} does not run again, since the host is already known and classified, whereas \emph{Dispatch}, \emph{Scan Preparation}, and \emph{Risk Assessment} are repeated. Aggregating the per-stage latencies in Table~\ref{tab:pipeline_latency}, a fast-only assessment introduces an orchestration overhead of 16.56~s (stages~1--3 and~5), which is is approximately 2.0\% of the 14-minute fast scan. A fast-to-deep assessment repeats stages~2, 3, and~5 for the deep pass, adding 10.04~s for a total of 26.60~s, which is approximately 0.73\% of the dual-stage scanning duration. In both cases, the orchestration overhead is always a small fraction of the scan-execution time, which spans tens of minutes (Tables~\ref{tab:port_reduction} and~\ref{tab:constrained_scan}), confirming that the framework's own processing overhead is not the bottleneck on the pipeline latency.

\begin{table}[htbp]
\caption{Decision-to-Mitigation Latency}
\label{tab:enforcement_latency}
\centering
\renewcommand{\arraystretch}{1.3}
\setlength{\tabcolsep}{4pt}
\begin{tabular}{clccc}
\hline
\textbf{\#} & \textbf{Action} & \textbf{Mean (ms)} & \textbf{Max (ms)} & \textbf{Std (ms)} \\
\hline
1 & Quarantine (host-wide)$^\dagger$ & 3.74 & 4.88 & 0.42 \\
2 & L4 drop (per-service)            & 3.66 & 6.28 & 0.77 \\
3 & SPAN mirror (per-service)        & 3.29 & 4.93 & 0.70 \\
\hline
\end{tabular}

\vspace{4pt}
{\footnotesize $^\dagger$\,Rule active on the Data Plane; the host retains its production address until it renews its DHCP lease.\par}
\end{table}

Table~\ref{tab:enforcement_latency} reports the elapsed time between the mitigation decision and the corresponding OpenFlow rule becoming active in the data plane. All three mitigation actions, quarantine, Layer-4 blocking, and traffic mirroring, become effective within a few milliseconds, several orders of magnitude faster than the preceding vulnerability assessment and risk evaluation stages. In addition, the low std values indicate that mitigation latency is consistently bounded, demonstrating predictable enforcement performance even under repeated executions.

Quarantine mitigation also depends on DHCP lease renewal. The VLAN reassignment and associated flow-rule updates are applied within milliseconds, immediately isolating the host from the production network, but the host cannot yet communicate within the quarantine VLAN until DHCP renewal completes. As reported in~\cite{11205517}, the average transition delay is approximately half the configured lease duration and can be significantly reduced by configuring shorter DHCP lease times for managed hosts.

Overall, the results demonstrate that the proposed framework introduces only modest orchestration overhead relative to the vulnerability assessment itself, while mitigation decisions are enforced within a few milliseconds. These findings confirm that the framework satisfies the latency requirements of automated vulnerability management and that its operational cost scales primarily with scan execution rather than with the orchestration pipeline.

\subsection{Risk Band Assignment Behavior}
\label{sec:risk_band_assignment}

\noindent This test addresses RQ2. It studies how CVSS severity, exploitation likelihood, and operational context are integrated into a dynamic risk assessment. It illustrates the effect of incorporating these three components sequentially: beginning with CVSS severity alone; then, adding exploitation evidence; and, finally applying context. The objective is to illustrate how each component influences the risk band assigned to a vulnerability using five representative vulnerabilities identified on a Metasploitable2 host. To the best of our knowledge, no comprehensive ground-truth dataset exists that accurately labels the exploitation status of arbitrary vulnerabilities. Consequently, this experiment illustrates the behavior of the proposed model rather than attempting to quantify its predictive accuracy. Table~\ref{tab:risk_bands} summarizes the test results. \par

\begin{table*}[t]
\centering
\caption{Risk Band Assignment Across CVSS-only and Model Configurations}
\label{tab:risk_bands}
\footnotesize
\setlength{\tabcolsep}{4pt}
\renewcommand{\arraystretch}{1.15}
\resizebox{\textwidth}{!}{
\begin{tabular}{p{2.2cm}p{1.6cm}ccccccc}
\hline
 & & & & & \multicolumn{4}{c}{\textbf{Assigned Risk Band (Model: @EPSS Cutoff; Normal / Sensitive Zone)}} \\
\textbf{CVE} & \textbf{Port} & \textbf{CVSS} & \textbf{EPSS} & \textbf{KEV} & \textbf{CVSS-only} & \textbf{Model @0.1 (Normal)} & \textbf{Model @0.1 (Sensitive)} & \textbf{Model @0.2 (Normal)} \\
\hline
CVE-2009-3099 & 8180/tcp & 10.0 & 0.8783 & no  & Critical & Critical & Critical & Critical \\
(no CVE)      & 8787/tcp & 10.0 & 0.0000 & no  & Critical & Critical & Critical & Critical \\
CVE-2020-1938 & 8009/tcp & 9.8  & 0.9447 & yes & Critical & Critical & Critical & Critical \\
CVE-1999-0678 & 80/tcp   & 5.0  & 0.1986 & no  & Medium   & High     & Critical & Low \\
CVE-2008-5304 & 80/tcp   & 10.0 & 0.0649 & no  & Critical & Medium   & High     & Medium \\
\hline
\end{tabular}}
\end{table*}

The model classifies exploitability into two levels: high and low. A finding is high when it is listed in KEV or when its EPSS probability is at or above a cutoff, and low otherwise. The cutoff therefore represents the EPSS probability at which a finding is treated as likely to be exploited. Its value is not authoritative: FIRST.org, which maintains EPSS, uses 0.1 only as an explicitly arbitrary example and states that it neither endorses nor opposes it. The threshold therefore represents an operational policy decision rather than a universal constant, which is why the framework exposes it as a configurable parameter. The value 0.1 is adopted here as a common reference point, close to the 0.088 threshold that Shimizu and Hashimoto derived empirically from a dataset of real vulnerabilities~\cite{shimizu26}. It was not calibrated against a labeled exploitation dataset in this work, and such calibration can be explored in future investigations.\par

The first step adds exploitation evidence to severity, comparing the CVSS-only column, which reflects the severity score, with the exploitability column, which incorporates observed or predicted vulnerability exploitation. For the three initial vulnerabilities in Table~\ref{tab:risk_bands}, the risk band remains Critical, but for different reasons, as following discussed. CVE-2009-3099 has a high EPSS, so exploitation is predicted to be likely. CVE-2020-1938 appears in KEV, so exploitation is confirmed. The finding on 8787/tcp has no CVE, so under the model worst-case rule it is assumed exploited. In each case, the model reaches Critical through evidence rather than by default, and aligns with severity. The last two vulnerabilities in Table~\ref{tab:risk_bands} are where the two disagree. CVE-1999-0678 scores only 5.0, but its EPSS exceeds the cutoff, so the model raises it from Medium to High. CVE-2008-5304 scores the maximum 10.0, but it has a low EPSS and it is absent from KEV, so the model lowers it from Critical to Medium. The model adjusts the initial severity-based prioritization in both directions, elevating vulnerabilities that are likely to be exploited while de-prioritizing severe vulnerabilities for which exploitation appears unlikely. \par

The second step uses operational context to refine the risk band. The same Metasploitable2 host is assessed first in a normal zone and then in a sensitive zone, so the only difference between the two columns is the host's exposure. In the sensitive zone, CVE-1999-0678 rises from High to Critical and CVE-2008-5304 from Medium to High. This risk band promotion does not alter the underlying vulnerability score; instead, it adjusts the assigned risk band to reflect the greater operational impact associated with a more exposed deployment environment. The three initial findings in Table~\ref{tab:risk_bands} do not change their risk bands, since Critical is already the topmost risk band. Exposure here is determined by the network zone, and the same promotion also applies to constrained devices, which does not affect the Metasploitable2 host.\par

The last column shows the effect of the EPSS cutoff, comparing exploitability at 0.1 with the same setup at 0.2. A higher cutoff requires a higher probability before a finding is classified exploited, so fewer findings meet the threshold. CVE-1999-0678 illustrates this case: its EPSS of 0.1986 is above 0.1 but below 0.2, so it is qualified as exploited under the lower cutoff but not under the higher cutoff, dropping from High to Low. The cutoff acts as a configurable risk-appetite parameter. A lower value triggers earlier intervention to a broader set of findings, while a higher value timely reacts only to the most probable high-risk vulnerabilities. The framework therefore keeps this cutoff configurable, allowing each organization to set it according to its own risk-tolerance and primary business goals.\par

The results presented in Table~\ref{tab:risk_bands} demonstrate how the proposed contextual risk model progressively refines vulnerability prioritization as additional information becomes available. While CVSS severity provides an initial estimate of technical impact, the incorporation of exploitation likelihood and operational context enables a more realistic assessment of the actual risk posed by each vulnerability within its deployment environment. Consequently, vulnerabilities with identical CVSS scores may be assigned different risk bands according to their probability of exploitation and their potential operational impact, supporting more proportionate mitigation decisions.

The examples analyzed in the current subsection demonstrate that the proposed model avoids treating CVSS severity as the sole determinant of mitigation priority, instead combining technical severity, exploitation evidence, and operational context to derive a risk band that more accurately reflects the expected operational risk.

While this case study demonstrates the qualitative behavior of the proposed model, it does not provide statistical evidence of its effectiveness at scale. Accordingly, the following subsection presents a statistical validation of the proposed context-aware prioritization approach using a case-control methodology based on vulnerabilities with verified exploitation evidence. The objective is to evaluate whether the proposed model can effectively reduce the number of vulnerabilities requiring immediate mitigation while maintaining complete coverage of known exploited vulnerabilities, thereby demonstrating its practical value for efficient vulnerability management in real-world scenarios.

\subsection{Exploitation-Aware Risk Prioritization for Efficient Vulnerability Management}
\label{sec:exploitation_aware_risk}

The previous subsection qualitatively demonstrated how the proposed risk model dynamically adjusts vulnerability risk bands by progressively incorporating CVSS severity, exploitation likelihood, and operational context. While those representative examples illustrate the expected behavior of the model, they do not quantify its effectiveness in supporting vulnerability management. Therefore, this subsection statistically validates the exploitation-aware component of the risk assessment model using a case-control methodology, which has been widely adopted in vulnerability research to evaluate the relationship between vulnerability characteristics and observed exploitation in the wild \cite{allodi14}. The objective is to evaluate whether the proposed model provides better discriminative capability than a conventional CVSS-only prioritization strategy by reducing unnecessary mitigation effort while maintaining complete coverage of vulnerabilities with verified exploitation. The contextual component is not exercised in this experiment, since the evaluation operates on published vulnerability data rather than on deployed hosts, and it therefore remains validated qualitatively in the previous subsection. The current test aims to address RQ2.

To isolate the contribution of exploitation likelihood risk assessment from intrinsic vulnerability severity, the evaluation considered only vulnerabilities published between January~1 and December~31,~2025 with a CVSS Base Score greater than or equal to 9.0. Restricting the analysis to vulnerabilities already classified as \emph{Critical} according to CVSS ensures that every vulnerability receives the same severity classification under the baseline approach, thereby eliminating severity as a confounding factor. Consequently, any observed improvement in prioritization can be attributed to the incorporation of exploitation evidence rather than to differences in vulnerability severity.

The ground truth for exploitation was established using the CISA Known Exploited Vulnerabilities (KEV) catalog. Within this dataset, KEV entries constitute the positive case (\emph{Cases}), whereas non-KEV vulnerabilities form the negative class (\emph{Controls}). Following standard recommendations for case-control studies, a random 1:4 matching strategy was adopted, selecting four control vulnerabilities for each exploited vulnerability to increase statistical power while minimizing sampling bias.

Each vulnerability was evaluated by the proposed contextual risk model, which combines CVSS severity with exploitation likelihood, derived from the EPSS exploit probability and KEV membership, to assign a risk band. The resulting prioritization was then compared against a conventional CVSS-only strategy, where all vulnerabilities with CVSS $\geq$ 9.0 were classified as requiring immediate mitigation. The evaluation focuses on the model's ability to distinguish exploited from non-exploited vulnerabilities within the highest CVSS severity class. 

\begin{figure}[hbt!]
\centering
\includegraphics[width=\columnwidth]{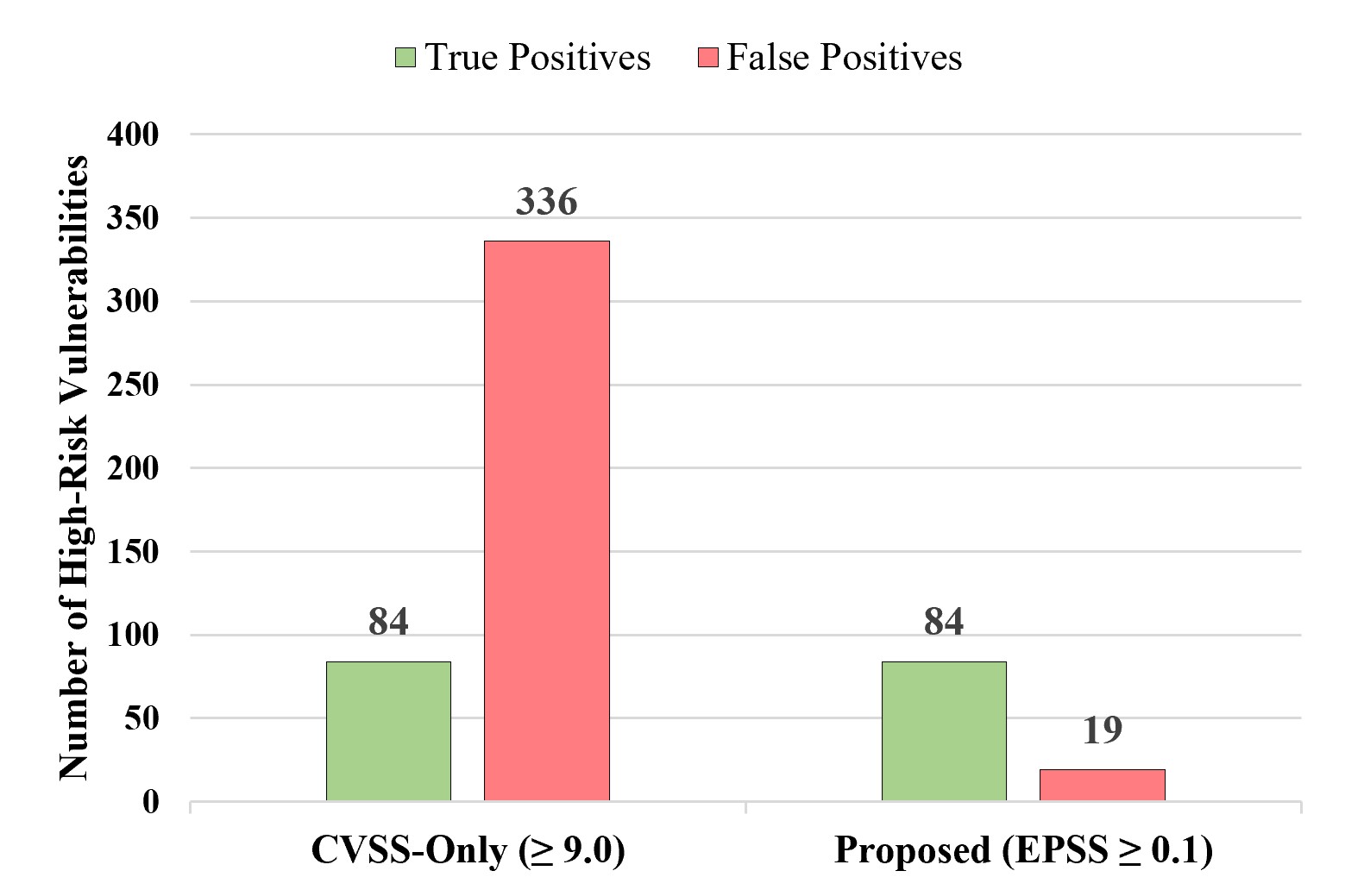}
\caption{Statistical validation of the proposed risk assessment model}
\label{fig:risk_model}
\end{figure}

For this evaluation, a \emph{True Positive (TP)} corresponds to a vulnerability listed in the CISA Known Exploited Vulnerabilities (KEV) catalog that is correctly prioritized for immediate mitigation. Conversely, a \emph{False Positive (FP)} corresponds to a vulnerability without verified exploitation that is nevertheless prioritized for immediate mitigation, unnecessarily increasing the mitigation workload. Since all exploited vulnerabilities were correctly identified by both the CVSS-only baseline and the proposed contextual risk model, no False Negatives (FNs) were observed in this experiment. Therefore, the evaluation primarily compares the reduction in False Positives while preserving complete detection of the True Positives. The obtained results are presented in Fig.~\ref{fig:risk_model}.

Figure~\ref{fig:risk_model} shows that both approaches correctly prioritized the 84 vulnerabilities with verified exploitation (green bars), thereby achieving complete coverage of the known exploited vulnerabilities. However, the CVSS-only strategy also classified an additional 336 non-exploited vulnerabilities as requiring immediate mitigation (red bar), highlighting the limited discriminative capability of severity-only prioritization. In contrast, the proposed model reduced the number of false positives from 336 to 19, a reduction of approximately 94\%, while preserving all true positives.

The 19 false positives should not be interpreted as model errors but rather as an expected consequence of incorporating predictive exploitation evidence into the prioritization process. The model raises a vulnerability to immediate mitigation when its predicted exploitation probability is high, whether or not exploitation has already been confirmed, so vulnerabilities with a high EPSS and no KEV entry stay in the priority set by design. Escalating on confirmed exploitation alone would drive the false-positive count to zero, but it would also remove any predictive capability and reduce the model to a retrospective reading of the ground truth. The remaining false positives are therefore the cost of anticipating exploitation rather than only reacting to it.

Overall, these results show that incorporating exploitation evidence significantly improves vulnerability prioritization over severity-only assessment. The proposed model reduces vulnerabilities requiring immediate remediation by approximately 75\% (420 to 103) while maintaining complete coverage of all verified exploited vulnerabilities, enabling security teams and automated response systems to focus on the highest-risk threats.

\subsection{Scalability}
\label{sec:scalability}

\noindent This test addresses RQ4 by evaluating the scalability of the proposed framework as the number of concurrently assessed hosts increases from 1 to 32 on a hybrid testbed comprising both physical and virtual machines. In operational deployments, host arrivals can be reasonably modeled by a stochastic process (e.g., a Poisson process), naturally distributing vulnerability assessments over time and reducing peak CPU, memory, network, and scanner utilization while preserving the same vulnerability management functionality. By contrast, the reported measurements represent a worst-case operational scenario, in which all hosts are scanned simultaneously, thereby maximizing resource utilization and providing an upper bound on the framework's execution time.

This experiment evaluates the vulnerability scanning subsystem in isolation rather than the complete framework because, as demonstrated in Section~\ref{sec:pipeline_latency}, scan execution is by far the dominant contributor to the framework's overall execution time. While a host assessment requires several minutes to complete, the remaining orchestration tasks, including host classification, risk assessment, and mitigation orchestration, complete within only a few seconds. Consequently, the scanner constitutes the scalability bottleneck of the framework, and stressing it directly provides an upper bound on the framework's scalability. Nevertheless, the scanner is executed using the framework's scan configurations, including the two-stage fast/deep progression and the class-specific scan profiles (reconnaissance-only for constrained devices and the full profile for traditional IT hosts). Therefore, the generated workload accurately reflects the operational behavior of the complete framework, with only the negligible orchestration overhead omitted.

\begin{figure}[hbt!]
\centering
\includegraphics[width=\columnwidth]{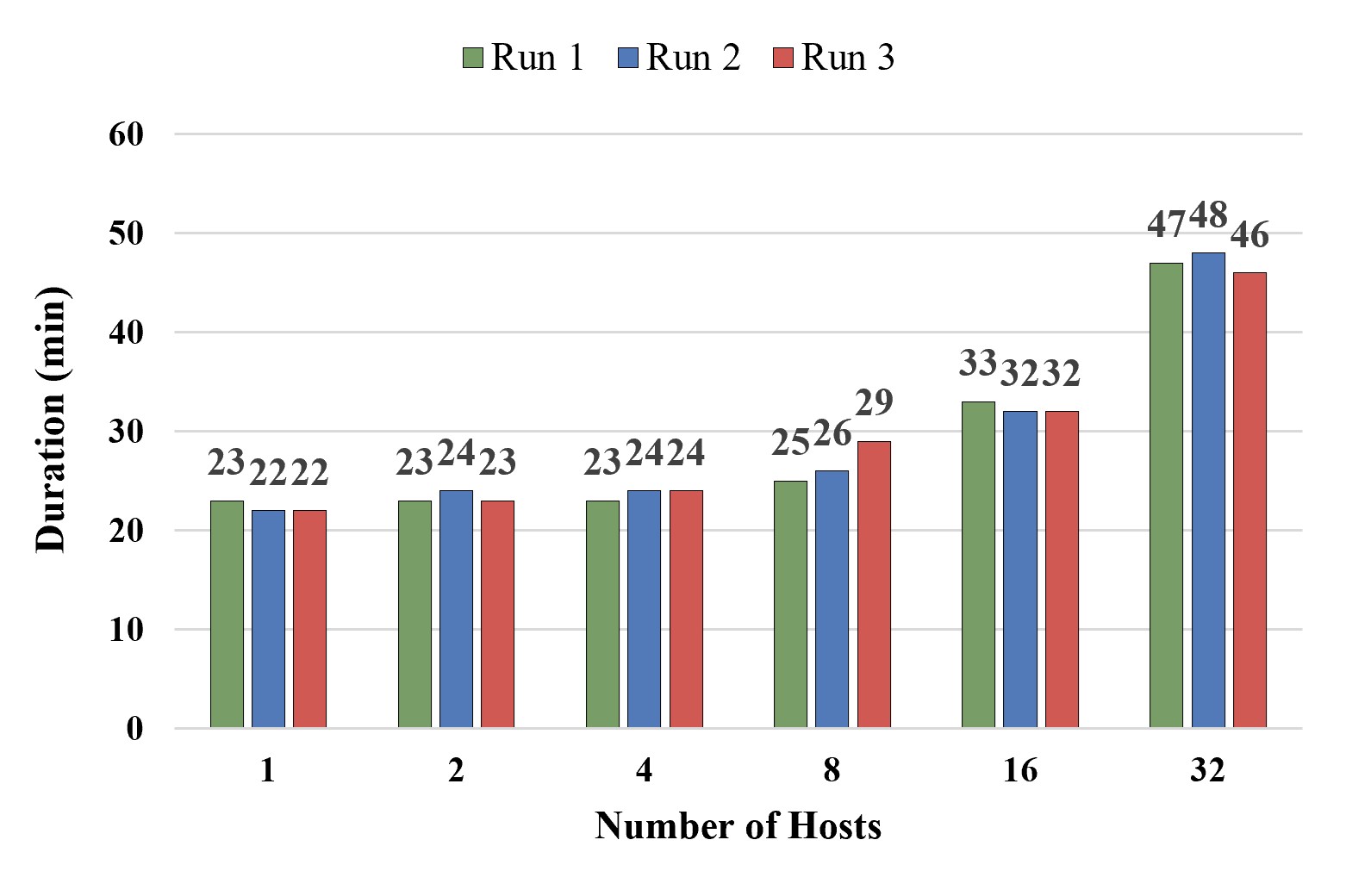}
\caption{Execution time (in minutes) required to assess 1-32 Metasploitable2 hosts}
\label{fig:scala_metasploitable}
\end{figure}

The results show that scan duration increases with the number of concurrently assessed hosts, but much more slowly than the increase in host count. As shown in Fig.~\ref{fig:scala_metasploitable} and Table~\ref{tab:scalability_meta}, the execution time of the traditional IT scan profile increases by only 2.2$\times$ as concurrency grows from 1 to 32 hosts. For each concurrency level, the reported statistics are aggregated over three runs: \textit{Max} is the highest CPU (or RAM) utilization observed at any point in time across all runs; \textit{Avg} is the mean of the per-run averages; and \textit{Std} is the standard deviation of those per-run averages, providing an indication of the consistency of resource utilization across repeated runs. Maximum CPU utilization reaches 100\% at some points during each scan while four or more hosts are analyzed in parallel, whereas memory usage remains low and nearly constant. These results indicate that the scalability of the vulnerability assessment process is primarily CPU-bound rather than memory-bound, suggesting that additional processing cores would yield greater performance improvements than increasing the available RAM. Future work will investigate distributed adaptive scanning that dynamically adjusts scan concurrency and distributes assessment tasks across multiple scanning engines based on available resources and estimated device risk.

\begin{table}[htbp]
\caption{Scalability under Concurrent Scans of Metasploitable2 Devices}
\label{tab:scalability_meta}
\centering
\renewcommand{\arraystretch}{1.3}
\setlength{\tabcolsep}{4pt}
\begin{tabular}{cccccccc}
\hline
\textbf{Hosts} & \textbf{Duration (min.)} & \multicolumn{3}{c}{\textbf{CPU (\%)}} & \multicolumn{3}{c}{\textbf{RAM (\%)}} \\
 & & \textbf{Max} & \textbf{Avg} & \textbf{Std} & \textbf{Max} & \textbf{Avg} & \textbf{Std} \\
\hline
1  & 22 & 75.5 & 17.4 & 0.8 & 13.3 & 12.3 & 0.5 \\
2  & 23 & 90.9 & 30.8 & 1.3 & 14.6 & 13.8 & 0.4 \\
4  & 24 & 100  & 52.9 & 0.8 & 15.5 & 14.1 & 1.1 \\
8  & 27 & 100  & 72.8 & 2.1 & 17.7 & 15.8 & 0.5 \\
16 & 32 & 100  & 87.5 & 1.9 & 22.0 & 19.4 & 0.5 \\
32 & 48 & 100  & 96.3 & 1.9  & 32.2 & 27.2 & 1.8 \\
\hline
\end{tabular}
\end{table}

\begin{figure}[hbt!]
\centering
\includegraphics[width=\columnwidth]{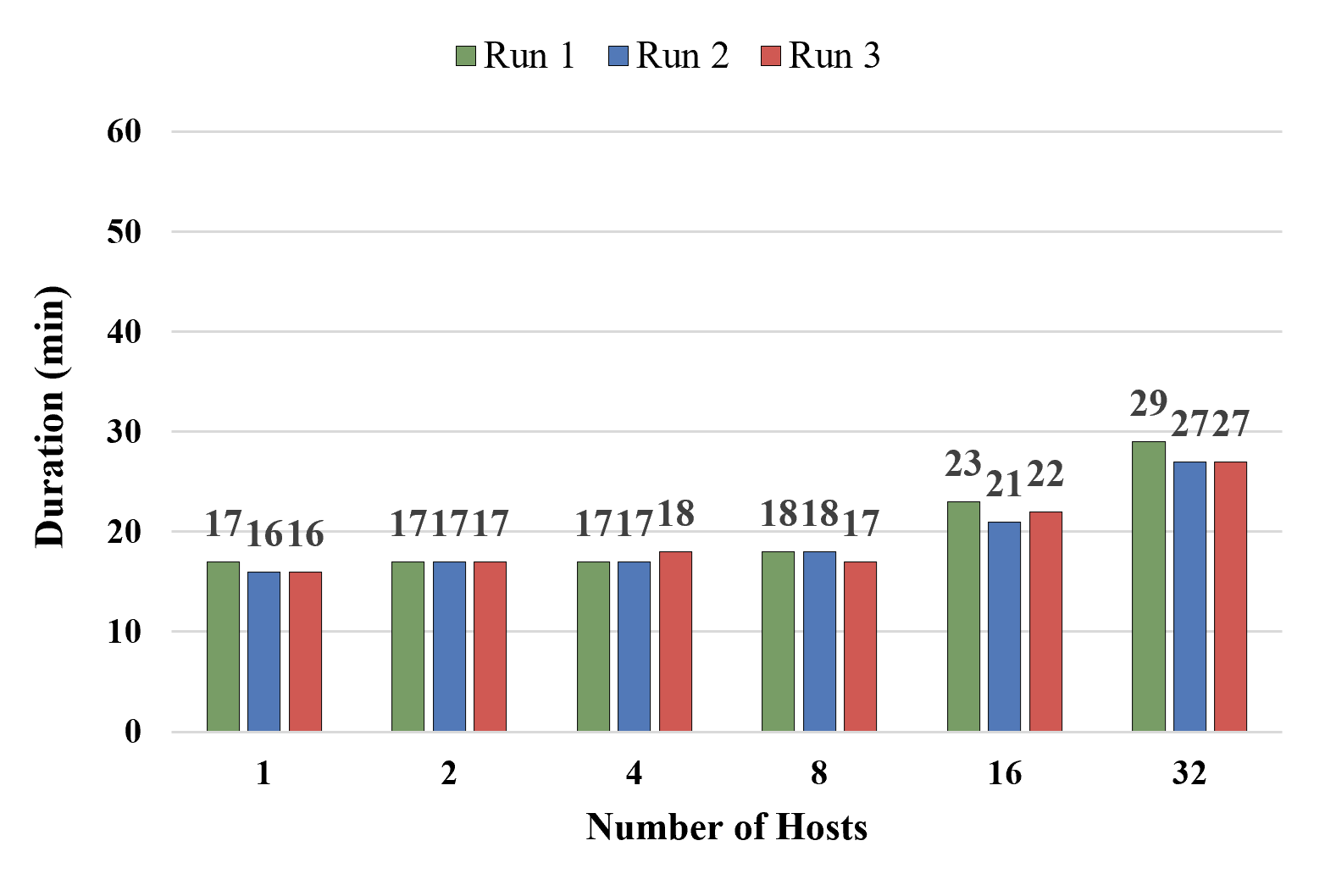}
\caption{Execution time (in minutes) required to assess 1-32 IoTGoat hosts.}
\label{fig:scala_iot}
\end{figure}

The scaling behavior also differs across device classes. The constrained-device profile scales more efficiently than the IT profile under concurrent workloads. As concurrency rises from 1 to 32 hosts, the constrained-device profile increases by only about 1.75$\times$ (Table~\ref{tab:scalability_iot}, Fig.~\ref{fig:scala_iot}), compared with approximately 2.2$\times$ for the traditional IT profile (Table~\ref{tab:scalability_meta}, Fig.~\ref{fig:scala_metasploitable}). This difference is explained by the reduced reconnaissance-only test set executed by the constrained-device profile. Consequently, each scan requires less processing and completes sooner, resulting in a smaller increase in execution time as the number of concurrent hosts grows. Although CPU utilization becomes high under both profiles as concurrency increases, the constrained-device profile performs substantially less work per host, resulting in shorter scan times and a slower increase in execution time. Therefore, the observed scalability difference is primarily due to the reduced scanning workload rather than differences in resource availability. The lighter treatment of fragile devices, introduced to preserve operational safety, also improves the scalability of the framework for this class of devices.

\begin{table}[htbp]
\caption{Scalability under Concurrent Scans of IoTGoat Devices}
\label{tab:scalability_iot}
\centering
\renewcommand{\arraystretch}{1.3}
\setlength{\tabcolsep}{4pt}
\begin{tabular}{cccccccc}
\hline
\textbf{Hosts} & \textbf{Duration (min.)} & \multicolumn{3}{c}{\textbf{CPU (\%)}} & \multicolumn{3}{c}{\textbf{RAM (\%)}} \\
 & & \textbf{Max} & \textbf{Avg} & \textbf{Std} & \textbf{Max} & \textbf{Avg} & \textbf{Std} \\
\hline
1  & 16 & 54.1 & 16.5 & 0.9  & 14.4 & 14.1 & 0.8  \\
2  & 17 & 95.4 & 31.0 & 0.8  & 15.3 & 14.7 & 0.6  \\
4  & 17 & 100  & 55.4 & 0.8  & 14.0 & 13.4 & 0.7  \\
8  & 18 & 100  & 86.3 & 2.1  & 18.6 & 17.5 & 0.3  \\
16 & 23 & 100  & 95.0 & 1.2  & 20.7 & 19.4 & 1.0  \\
32 & 28 & 100  & 97.5 & 1.4  & 28.2 & 26.5 & 1.5 \\
\hline
\end{tabular}
\end{table}

\begin{figure}[hbt!]
\centering
\includegraphics[width=\columnwidth]{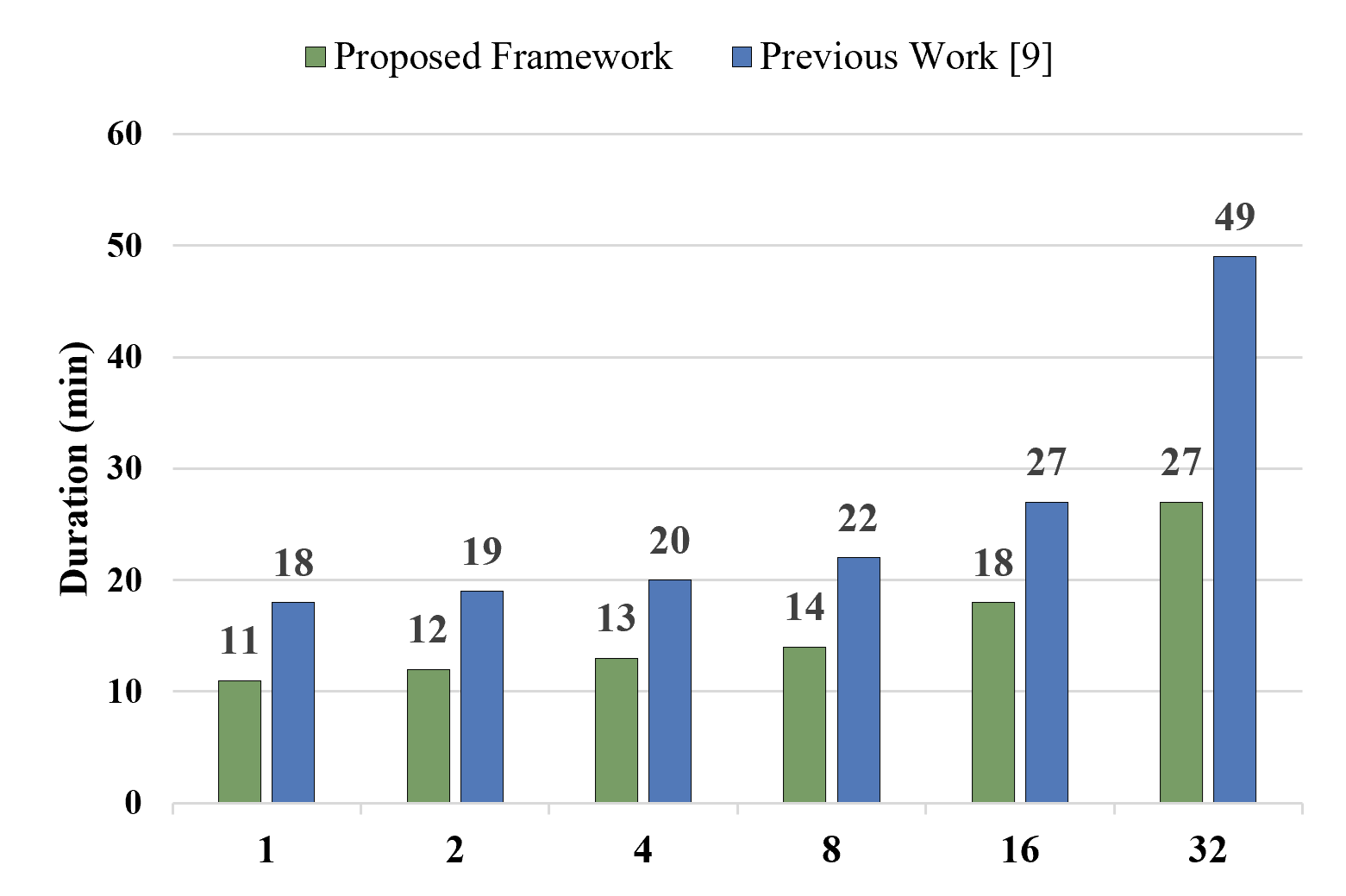}
\caption{Time (in minutes) to assess 1--32 Windows~2003 hosts for the proposed framework against the baseline approach~\cite{11205517}.}
\label{fig:scala_win_comparison}
\end{figure}

To contextualize these results against previous work, the scalability of the proposed framework was compared with the OpenVAS-based vulnerability assessment framework presented in~\cite{11205517}, evaluated under the same physical testbed environment. The evaluation measures the execution time required as the number of concurrently scanned physical hosts increases from 1 to 32 (Fig.~\ref{fig:scala_win_comparison}). Notably, the experimental configurations feature unequal resource allocations: the proposed framework operates using a single CPU core, whereas the baseline approach in~\cite{11205517} utilizes eight CPU cores. Despite this substantially more constrained computational environment, the proposed framework consistently achieves shorter assessment times across the entire evaluated range. The largest performance advantage is observed at peak load, where the proposed framework assesses 32 concurrent hosts in 27~minutes, the same time the baseline approach requires to assess only 16 hosts. Consequently, the proposed solution approximately doubles the assessment throughput within a comparable time budget. The scaling behavior further reflects this advantage: as host concurrency increases from 1 to 32, the execution time of the baseline approach grows by approximately $2.7\times$ (from 18 to 49~minutes), whereas the proposed framework increases by only about $2.4\times$ (from 11 to 27~minutes).

This improvement results from the framework's adaptive scanning strategy, which combines threat-informed port prioritization, targeted vulnerability-test selection, device-specific scan profiles, and a two-stage fast/deep assessment process. Rather than performing an exhaustive assessment before taking action, the framework first executes a lightweight fast scan that focuses on the most frequently attacked ports and the most relevant vulnerability tests. As demonstrated in Section~\ref{sec:constrained_aware_scanning} (Table~\ref{tab:constrained_scan}), dividing the assessment into fast and deep stages introduces virtually no additional execution time, since the combined duration is approximately the same as that of a single full-range scan. More importantly, the fast stage enables the framework to establish an initial security posture and initiate risk-informed mitigation actions well before the deep assessment has completed, thereby reducing the exposure window while preserving the comprehensive coverage provided by the subsequent scan.

Overall, the scalability evaluation demonstrates that the proposed framework can efficiently support concurrent vulnerability assessments while maintaining low memory consumption. The results identify CPU utilization within the scanning component as the primary scalability bottleneck, indicating that future work should focus on improving assessment throughput through adaptive scan scheduling, distributed scanning architectures, and dynamic load balancing across multiple scanning engines. In addition, incorporating risk-aware scheduling policies that prioritize hosts according to their preliminary risk estimate could further optimize resource utilization by ensuring that computational resources are preferentially allocated to the most security-critical assets.

The next section analyzes the main contributions of this work, highlighting its strengths and limitations.  \par

\section{Lessons Learned}
\label{sec:discussion}

This section summarizes the main lessons learned from the proposed framework and its experimental evaluation. First, Section~\ref{sec:comparison_soa} positions the proposed framework with respect to the state of the art, emphasizing its distinguishing characteristics and the experimental evidence supporting them. Next, Section~\ref{sec:strengths_proposal} analyzes the principal strengths demonstrated throughout the proposal evaluation. Finally, Section~\ref{sec:limitations_proposal} identifies the current limitations of the framework and outlines promising directions for future improvements, providing a balanced assessment of the proposed approach.

\subsection{Comparison with State-of-the-Art}
\label{sec:comparison_soa}

The evaluation in Section~\ref{sec:evaluation} shows that, unlike existing solutions that address isolated stages of the vulnerability-management lifecycle, the proposed framework integrates adaptive discovery, context-aware risk assessment, and fine-grained automated mitigation into a unified end-to-end workflow.

Conventional vulnerability scanners generally apply the same scan profile to every asset. In contrast, the proposed adaptive mechanism adjusts the assessment depth according to the detected device class. By combining passive asset identification with class-specific active scanning profiles, the framework substantially reduces unnecessary scanning activity on resource-constrained devices while preserving the information required for accurate risk assessment. The experimental evaluation in Subsection~\ref{sec:constrained_aware_scanning}, Table~\ref{tab:constrained_scan}, demonstrated that this strategy reduces scan time by up to 91\% while maintaining 71\% vulnerability coverage after the initial fast scan. These results demonstrate that the proposed two-stage approach substantially accelerates vulnerability assessment while preserving sufficient coverage to support timely risk-informed mitigation. Nevertheless, the constrained profile cannot detect vulnerabilities that require active verification, including one critical backdoor in the evaluated testbed. The framework addresses this limitation through continuous passive monitoring, which complements the reconnaissance-only assessment by observing subsequent network activity. Quantitatively evaluating the effectiveness of this complementary monitoring strategy remains an important direction for future work.

The proposed context-aware risk assessment also provides a more operationally meaningful prioritization than traditional CVSS-only approaches. By integrating CVSS severity, EPSS exploitation probability, and operational context, the framework reduces by approximately 75\% the number of vulnerabilities requiring immediate mitigation while preserving complete coverage of vulnerabilities with verified exploitation. This enables security teams to focus mitigation efforts on vulnerabilities that represent the highest operational risk instead of treating every critical-severity vulnerability with the same priority.

Another distinguishing characteristic of the proposed framework is the integration of SDN and SOAR technologies to automate proportional mitigation actions. Rather than relying exclusively on disruptive host isolation, the framework dynamically selects mitigation actions ranging from traffic mirroring and monitoring to selective service isolation or complete host quarantine according to the assessed risk band. The experimental evaluation further demonstrated that these mitigation actions are enforced within milliseconds after the risk assessment completes, significantly reducing the exposure window between vulnerability identification and response. For host quarantine, the OpenFlow rules and VLAN reassignment are applied immediately; however, the host becomes fully operational within the quarantine network only after renewing its DHCP lease. Consequently, the effective isolation is immediate, whereas the completion of the network reconfiguration depends on the remaining DHCP lease duration.

Finally, the scalability evaluation demonstrates that the proposed architecture scales efficiently as the number of concurrently assessed hosts increases. Although vulnerability scanning remains the dominant computational cost, the execution time grows considerably more slowly than the assessment workload, achieving up to a 45\% lower execution time for the concurrent assessment of 32 hosts (Subsection~\ref{sec:scalability}) than a baseline OpenVAS-based vulnerability assessment approach \cite{11205517}. Moreover, whereas most vulnerability-management proposals are validated exclusively through simulation or fully virtualized environments, the proposed framework was experimentally validated on both virtualized and hybrid testbeds comprising physical and virtual devices, contributing experimental evidence to a research area that remains comparatively underexplored in the literature~\cite{souto2026moving}.

\subsection{Strengths of the Proposed Approach}
\label{sec:strengths_proposal}

The conducted evaluation demonstrates that the proposed framework successfully combines adaptive vulnerability discovery, context-aware risk assessment, and automated mitigation into a coherent end-to-end vulnerability management workflow.

First, the adaptive two-stage scanning strategy effectively balances assessment quality against operational latency. By combining passive identification with class-specific active scanning profiles, the framework safely assesses heterogeneous devices while significantly reducing unnecessary probing of resource-constrained IoT assets. The experimental results show that this strategy substantially reduces scan time while preserving the vulnerability coverage required for accurate risk assessment.

Second, the proposed context-aware risk model provides a more operationally meaningful vulnerability prioritization than conventional severity-only approaches. By integrating CVSS severity, EPSS exploitation probability, and operational context, the framework identifies a considerably smaller subset of vulnerabilities requiring immediate mitigation without sacrificing protection against vulnerabilities with verified exploitation. Consequently, both automated mitigation actions and security teams can focus their efforts on the vulnerabilities that contribute most to the overall operational risk.

Third, SOAR-SDN orchestration provides timely, proportional mitigation. Instead of relying exclusively on disruptive host isolation, the framework dynamically selects mitigation actions ranging from traffic mirroring and monitoring to service isolation or complete host quarantine according to the assigned risk band. The experimental evaluation further demonstrates that these actions are enforced within milliseconds after the risk assessment completes, considerably reducing the exposure window.

Finally, the scalability evaluation demonstrates that the proposed architecture scales predictably under increasing assessment workloads (range 1-32 simultaneous hosts). Although vulnerability scanning remains the dominant computational cost, the execution time increases substantially more slowly than the number of concurrently assessed hosts, indicating good scalability characteristics. Since the scalability bottleneck is localized within the vulnerability scanning component and is primarily caused by high CPU utilization, larger deployments can be effectively supported by provisioning additional processing resources or distributing the scanning workload across multiple scanning engines, without requiring changes to the overall framework architecture.

 \subsection{Limitations of the Proposed Approach}
\label{sec:limitations_proposal}

Despite the encouraging experimental results, several limitations should be acknowledged.

First, the proposed framework relies on the vulnerability coverage provided by the underlying scanning engine. Vulnerabilities that cannot be detected by the scanner cannot subsequently be prioritized or mitigated by the framework. Although the adaptive scanning strategy reduces assessment time, it cannot compensate for limitations in the scanner's vulnerability knowledge base. 

Second, the framework operates under the assumption of accurate device classification. The current architecture does not account for edge-case misclassifications, such as identifying a resource-constrained IoT device as a traditional IT host. This could inadvertently subject fragile assets to disruptive active probing. Evaluating the reliability of the classifier and implementing fail-safe mechanisms during the active scanning phase remain areas for future investigation.

Third, the contextual risk model depends on the quality and timeliness of external threat-intelligence sources, including EPSS, the CISA Known Exploited Vulnerabilities (KEV) catalogue, and attack telemetry feeds. Delayed, incomplete, or inaccurate threat intelligence may temporarily affect the prioritization assigned to newly disclosed vulnerabilities. Furthermore, the model's efficacy is bounded by a static, policy-driven EPSS threshold; dynamic calibration of this threshold against labeled real-world exploitation datasets is deferred to future iterations. In addition, many real-world exploitations are never publicly disclosed. To alleviate the limitations of this paragraph and last one, future versions of the framework can incorporate AI/ML models that timely correlate network telemetry, configuration changes, system logs, and historical attack patterns to identify anomalous behaviors and previously unknown security weaknesses, initiating their risk classification and posterior proper mitigation before exploitation.

Fourth, the mitigation is supported by a single centralized SDN controller, which can raise concerns of lack of availability and response time. A more robust control of mitigation actions can be deployed by a dynamic set of multiple coordinated SDN controllers.

Finally, the scalability evaluation identifies CPU resources within the vulnerability scanning component as the principal performance bottleneck. Although the framework scales well for the evaluated workload, substantially larger deployments would benefit from distributed scanning architectures, adaptive scan scheduling, and dynamic load balancing across multiple scanning engines. These directions constitute promising opportunities for extending the framework while preserving its modular architecture.

The next section concludes this paper by summarizing its main contributions and outlining future research directions.

\section{Conclusions and Future Work}
\label{sec:conclusions}

This section summarizes the scientific and practical contributions of the proposed framework. Section~\ref{sec:summary_contributions} reviews the main outcomes and their contribution to automated vulnerability management in programmable heterogeneous networks. Section~\ref{sec:future_research} then outlines research directions for improving the architecture and extending its applicability to future network environments.

\subsection{Summary of Contributions}
\label{sec:summary_contributions}

This paper proposed a unified framework for managing vulnerabilities in heterogeneous programmable networks. The framework combines adaptive vulnerability discovery, context-aware risk assessment, and SDN-based mitigation through a SOAR pipeline. Conventional solutions typically address scanning, prioritization, and mitigation independently. In contrast, the proposed architecture coordinates these stages to provide continuous, context-driven vulnerability management.

The experimental evaluation demonstrated that the proposed framework successfully addresses the four research questions formulated in this work. The adaptive two-stage vulnerability assessment strategy significantly reduces unnecessary scanning overhead by tailoring scan depth to the capabilities of each device class while preserving vulnerability coverage. The context-aware risk assessment model combines CVSS severity, exploitation likelihood, and operational context to prioritize vulnerabilities more effectively than conventional severity-only approaches. It reduces the number of vulnerabilities requiring immediate mitigation while maintaining complete coverage of vulnerabilities with verified exploitation. Furthermore, the SDN-SOAR integration enables mitigation actions to be automatically enforced within milliseconds after risk assessment, thereby minimizing the exposure window between vulnerability identification and response. Finally, the scalability evaluation showed that the framework scales efficiently under increasing assessment workloads, with CPU resources constituting the primary scalability constraint.

Overall, the obtained results demonstrate that combining adaptive two-stage scanning, contextual risk assessment, and automated mitigation enables a more efficient vulnerability management process than treating these functions independently. By reducing unnecessary scanning, prioritizing only the vulnerabilities that present the greatest operational risk, and automating the corresponding mitigation actions, the proposed framework decreases both the operational workload imposed on security teams and the time required to automatically reduce the attack surface of heterogeneous programmable networks.

\subsection{Future Research}
\label{sec:future_research}

Although the proposed framework demonstrates promising results, several research directions remain.

First, artificial intelligence can further enhance the vulnerability management pipeline \cite{zeng2021licality,11527394}. Machine learning can improve exploitation prediction, dynamic risk assessment, and scan scheduling, while reinforcement learning can optimize mitigation policies by balancing security, service availability, and operational impact.

Second, the mitigation capabilities can be extended beyond SDN-based enforcement to include Moving Target Defense (MTD)~\cite{souto2026moving}, micro-segmentation, adaptive access control, deception technologies, and dynamic service migration. In addition, distributed SDN controllers~\cite{10252983} can improve scalability, resilience, and fault tolerance in large-scale deployments.

Third, integrating the framework with Zero Trust architectures \cite{10.1016/j.adhoc.2024.103414} would enable continuous, risk-aware access control by incorporating vulnerability exposure, exploitability, and operational context into trust decisions.

Finally, future work will investigate distributed scanning architectures, adaptive scan orchestration, contactless active reconnaissance \cite{8688018}, and coordinated scanning engines to overcome scalability limitations and support larger deployments while preserving the framework's automation and responsiveness.

\bibliographystyle{IEEEtran}
\bibliography{refs}

\begin{IEEEbiography}[{\includegraphics[width=1.0in,height=1.25in,clip,keepaspectratio]{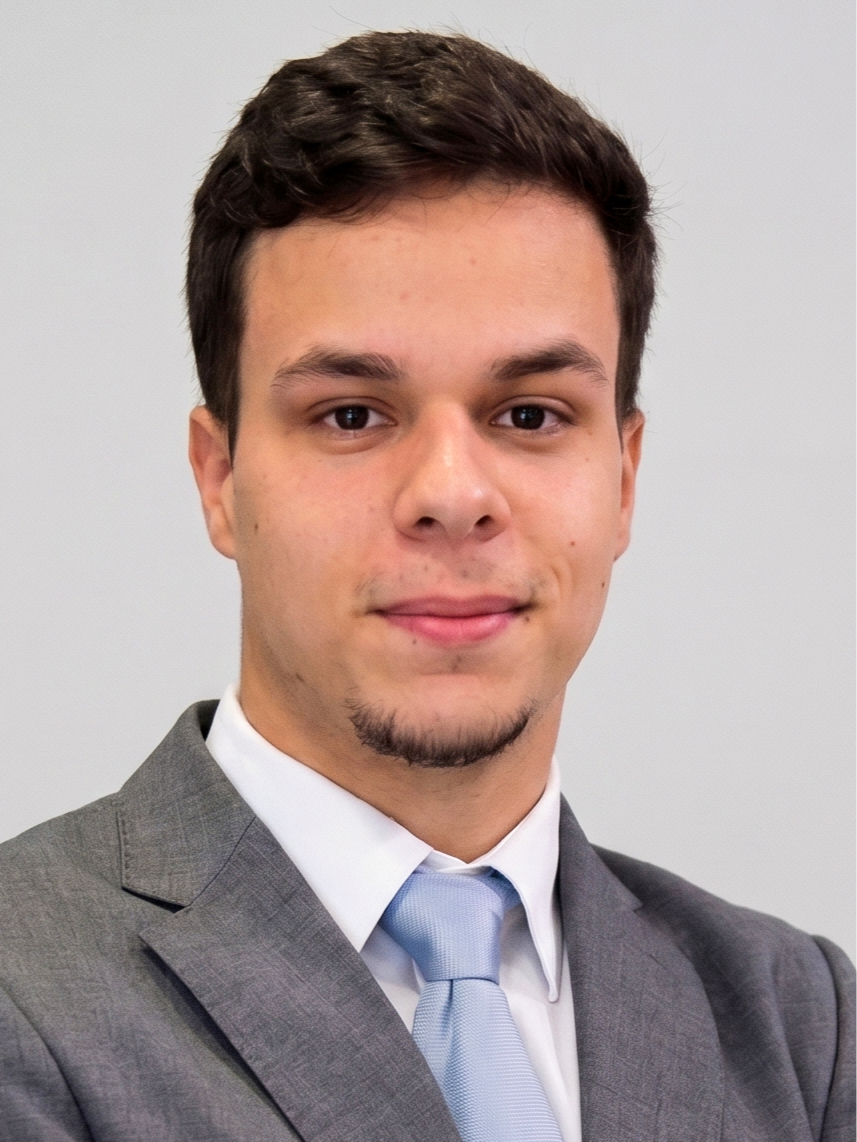}}]{Ricardo Lopes} received the B.Sc. degree in Telecommunications and Computer Engineering from Iscte - Instituto Universitário de Lisboa, Portugal, in 2024, and the M.Sc. degree in the same field from Iscte in 2026. His research interests include computer networks and network security, which he developed throughout his undergraduate and graduate studies. During his time at Iscte, he has also contributed to teaching activities in computer network architectures.
\end{IEEEbiography}

\begin{IEEEbiography}[{\includegraphics[width=1.0in,height=1.25in,clip,keepaspectratio]{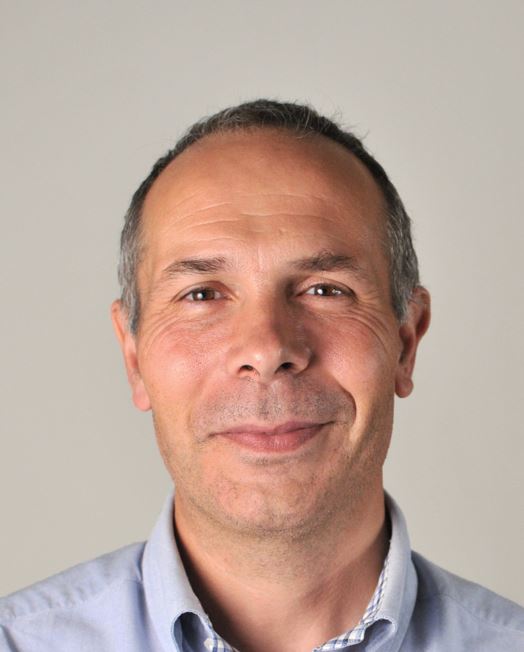}}]{José Moura} received the B.Sc. degree from the Universidade de Aveiro, Portugal, in 1989, the M.Sc. degree from the Universidade do Porto, Portugal, in 2001, and the Ph.D. degree from Lancaster University, U.K., in 2011. Since 2001, he has been with Iscte – Instituto Universitário de Lisboa, teaching Computer Networks. He is also a researcher with the Instituto de Telecomunicações, investigating network management, edge computing, optimization, software-defined networking, security, and resilient networked systems.
\end{IEEEbiography}

\begin{IEEEbiography}[{\includegraphics[width=1.0in,height=1.25in,clip,keepaspectratio]{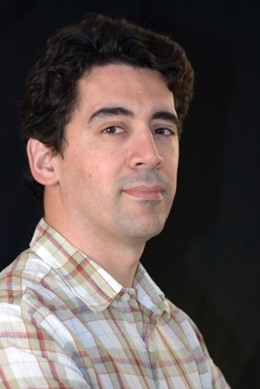}}]{Rui Neto Marinheiro} is an Associate Professor at Iscte – Instituto Universitário de Lisboa, Portugal, and a researcher at the Instituto de Telecomunicações. He received the Ph.D. degree in Multimedia Information Systems from the University of Southampton, U.K., and the M.Eng. degree in Electrical and Computer Engineering from the University of Porto, Portugal. His research interests include telecommunications, computer networks, security, and the Internet of Things. He has participated in national and international research projects. 
\end{IEEEbiography}

\end{document}